\documentclass[a4paper, 10pt]{article}

\usepackage{jheppub}

\usepackage{amsthm}

\usepackage[numbers,sort&compress]{natbib}
\usepackage{dsfont}
\usepackage{slashed}
\usepackage{color}
\usepackage{empheq}

\usepackage{pifont}

\usepackage{adjustbox}

\usepackage{bibgerm}
\usepackage{soul}

\usepackage[english]{babel}
\usepackage[utf8x]{inputenc}

\usepackage{graphicx,epsfig}
\usepackage{pstricks}

\usepackage{bashful}

\usepackage{subfigure}

\usepackage{setspace}

\usepackage{booktabs}

\usepackage{float}

\usepackage{nextpage}

\usepackage{pdfpages}

\usepackage{hypcap}

\usepackage[small,bf]{caption}

\usepackage{longtable}

\usepackage[makeroom]{cancel}

\usepackage{microtype}

\usepackage{xparse}
\usepackage{etoolbox}

\usepackage[subfigure]{tocloft}

\usepackage{listings}

\usepackage{enumitem}

\newcommand{\p}{\partial}
\newcommand{\pl}{\overleftarrow{\partial}\hspace{-0.75mm}}

\newcommand{\<}{\langle}
\renewcommand{\>}{\rangle}
\renewcommand{\O}{\mathcal{O}}
\newcommand{\tr}{\mathrm{tr}}

\renewcommand{\L}{\mathcal{L}}
\newcommand{\Q}{\mathcal{Q}}
\newcommand{\E}{\mathcal{E}}
\newcommand{\nn}{\nonumber\\}

\newcommand{\y}{\mathsf{y}}

\newcommand{\msbar}{$\overline{\text{MS}}$}

\newcommand{\hc}{\mathrm{h.c.}}

\newcommand{\GeV}{\,\text{GeV}}

\makeatletter
  \def\my@tag@font{\normalsize}
  \def\maketag@@@#1{\hbox{\m@th\normalfont\my@tag@font#1}}
  \let\amsmath@eqref\eqref
  \renewcommand\eqref[1]{{\let\my@tag@font\relax\amsmath@eqref{#1}}}
\makeatother

\makeatletter
\renewcommand\paragraph{\@startsection{paragraph}{4}{\z@}%
  {-3.25ex\@plus -1ex \@minus -.2ex}%
  {1.5ex \@plus .2ex}%
  {\normalfont\normalsize\bfseries}}
\makeatother

\cftsetindents{section}{0em}{2em}
\cftsetindents{subsection}{2em}{3em}
\cftsetindents{subsubsection}{5em}{4em}

\allowdisplaybreaks[1]

\preprint{ZU-TH 61/25}

\title{\boldmath Two-loop anomalous dimensions for baryon-number-violating operators in SMEFT}

\author[a,b]{Sumit Banik,}

\author[a]{Andreas Crivellin,}

\author[a,b]{Luca Naterop,}

\author[a,b]{Peter Stoffer}

\emailAdd{sumit.banik@physik.uzh.ch}
\emailAdd{andreas.crivellin@physik.uzh.ch}
\emailAdd{luca.naterop@physik.uzh.ch}
\emailAdd{stoffer@physik.uzh.ch}

\affiliation[a]{Physik-Institut, Universit\"at Z\"urich, Winterthurerstrasse 190, 8057 Z\"urich, Switzerland}
\affiliation[b]{PSI Center for Neutron and Muon Sciences, 5232 Villigen PSI, Switzerland}

\abstract{
We compute the two-loop renormalization-group equations for the baryon-number-violating dimension-six operators in the SMEFT. This includes all three gauge interactions, the Yukawa, and Higgs self-interaction contributions. In addition, we present the one-loop matching of the $S_1$ scalar leptoquark on the SMEFT, which can generate the Wilson coefficients of all four gauge-invariant baryon-number-violating SMEFT operators. Using this example, we demonstrate the cancellation of scheme and matching-scale dependences. Together with the known two-loop renormalization-group evolution below the electroweak scale in the LEFT, as well as the one-loop matching of SMEFT onto LEFT, our results enable consistent next-to-leading-log analyses of nucleon decays, provided that the relevant matrix elements are known at next-to-leading-order accuracy.
}

\numberwithin{equation}{section}

\begin{document}

	\maketitle



\section{Introduction}

Baryon number ($B$), defined as the number of baryons (such as protons and neutrons) minus the number of antibaryons (i.e., $1/3$ for quarks and $-1/3$ for anti-quarks), is conserved in all perturbative interactions within the Standard Model (SM). While non-perturbative sphaleron effects violate baryon number, the corresponding rates are undebatably small. This means that any observation of baryon-number violation (BNV) would be a clear indication of beyond-the-SM (BSM) physics. 

One primary motivation for BNV arises from the observed matter--antimatter asymmetry in the universe, which can be explained if the three Sakharov conditions, $C$ and $CP$ violation, departure from thermal equilibrium, and BNV are met~\cite{Sakharov:1967dj}. However, the sphaleron effects are at high temperatures insufficient to account for the observed asymmetry~\cite{Fukugita:1986hr}. Furthermore, new physics models, such as Grand Unified Theories (GUTs) or $R$-parity violating supersymmetry, predict or at least allow BNV through processes like proton decay~\cite{Georgi:1974sy,Langacker:1980js,Barbier:2004ez}, possibly at rates accessible with future experiments like Super-Kamiokande~\cite{Super-Kamiokande:2002weg} or DUNE~\cite{DUNE:2020ypp}.

In these cases, BNV usually occurs at very high energies so that its effects are strongly suppressed at low energies, ensuring a sufficiently long proton lifetime to satisfy the experimental constraints. In such a setup, the BNV effects can be parametrized within effective field theories (EFTs) via higher-dimension operators, starting at dimension~6. As the BSM scale is usually much higher than the electroweak scale, one can use the SM Effective Field Theory (SMEFT) with explicit $SU(2)_L$ invariance to reduce the number of operators~\cite{Buchmuller:1985jz,Grzadkowski:2010es}, see Refs.~\cite{Crivellin:2023ter,Beneito:2023xbk,Gargalionis:2024nij,Beneke:2024hox,Gisbert:2024sjw,Broussard:2025opd} for recent work. For the evolution of the Wilson coefficients below the electroweak scale, the SMEFT should be matched to the low-energy EFT (LEFT). This EFT approach not only simplifies the calculations but also improves the precision since the potentially huge scale separation between the BNV scale and the hadronic scale mandates the resummation of large logarithms.

The one-loop renormalization-group equations (RGEs) of these two EFTs at dimension~6~\cite{Jenkins:2013zja,Jenkins:2013wua,Alonso:2013hga,Alonso:2014zka,Jenkins:2017dyc}, together with the tree-level matching of the SMEFT on LEFT~\cite{Jenkins:2017jig} can be used for computations at leading-logarithmic (LL) accuracy. For next-to-leading-log (NLL) accuracy the one-loop matching between SMEFT and LEFT~\cite{Dekens:2019ept}, the two-loop RGEs~\cite{Buras:1989xd,Buras:1991jm,Buras:1992tc,Ciuchini:1993vr,Ciuchini:1993fk,Buchalla:1995vs,Chetyrkin:1997gb,Buras:2000if,Gorbahn:2016uoy,Panico:2018hal,deVries:2019nsu,Bern:2020ikv,Aebischer:2022anv,Fuentes-Martin:2022vvu,Aebischer:2023djt,Jenkins:2023rtg,Jenkins:2023bls,Naterop:2023dek,DiNoi:2023ygk,Fuentes-Martin:2023ljp,Morell:2024aml,Aebischer:2024xnf,Manohar:2024xbh,DiNoi:2024ajj,Born:2024mgz,Naterop:2024cfx,Fuentes-Martin:2024agf,Aebischer:2025hsx,Duhr:2025zqw,Haisch:2025lvd,Zhang:2025ywe,Naterop:2025lzc,Naterop:2025cwg,Duhr:2025yor,Haisch:2025vqj}, as well as the calculation of all observables at next-to-leading order (NLO) are necessary. Furthermore, the one-loop matching of UV models on the SMEFT at NLO is needed, which has been automated to a large extent~\cite{Carmona:2021xtq,Fuentes-Martin:2022jrf,Fuentes-Martin:2023ljp,Guedes:2023azv,Aebischer:2023nnv,Thomsen:2024abg,Guedes:2024vuf}.

For the BNV sector at dimension~6, the one-loop RGEs were calculated in Refs.~\cite{Abbott:1980zj,Alonso:2014zka,Jenkins:2017dyc}. Two- and three-loop QCD corrections were obtained in Refs.~\cite{Nihei:1994tx,Gracey:2012gx}. In the LEFT, the two-loop RGEs are already available in multiple schemes~\cite{Aebischer:2025hsx,Naterop:2025lzc}.\footnote{The disagreement between these two references is expected to disappear with an upcoming revision of Ref.~\cite{Aebischer:2025hsx}. In the following, we rely on the results of Ref.~\cite{Naterop:2025lzc}.} 
In the present article, we complete the two-loop RGE calculation for the BNV SMEFT operators at dimension~6. In addition to $SU(3)_c$ and $U(1)_Y$ gauge contributions as in the LEFT, our calculation includes the $SU(2)_L$ gauge and Yukawa interactions, as well as mixed gauge--Yukawa two-loop contributions and the Higgs self-interactions. In addition, we calculate the one-loop correction to the matching on SMEFT of the $S_1$ leptoquark, which can generate the Wilson coefficients of all four gauge-invariant operators. Therefore, our work completes the NLL EFT framework for the BNV sector at dimension~6, enabling refined analyses of the constraints from proton decay on heavy BNV new physics if combined with the NLO matching corrections~\cite{Gracey:2012gx} to non-perturbative determinations of the hadronic matrix elements~\cite{Yoo:2021gql}.

The rest of the article is structured as follows. In Sect.~\ref{sec:SMEFT}, we define our conventions for SMEFT, including the BNV operators and their flavor symmetries. We define our regularization and renormalization scheme in Sect.~\ref{sec:Scheme}, including the basis of evanescent operators. We discuss the renormalization procedure at one and two loops in Sect.~\ref{sec:Renormalization} and the results for the RGEs in Sect.~\ref{sec:Results}. In Sect.~\ref{sec:Matching}, we perform the one-loop matching of a leptoquark model to the SMEFT and we discuss the impact of NLL corrections, before concluding in Sect.~\ref{sec:Conclusions}. The full results for the two-loop RGEs and the one-loop matching are provided in electronic form as supplementary material, which uses the conventions explained in App.~\ref{sec:Conventions}.


\section{SMEFT}
\label{sec:SMEFT}

The SMEFT is the linearly realized EFT that is based on the SM gauge group $SU(3)_c \times SU(2)_L \times U(1)_Y$ and includes a tower of higher-dimension operators,
\begin{equation}
	\label{eq:SMEFT}
	\L_\mathrm{SMEFT} = \L_\mathrm{SM} + \sum_i C_i \Q_i + \sum_i K_i \E_i \, ,
\end{equation}
where $\Q_i$ are physical operators and $\E_i$ evanescent ones that vanish in $D=4$ space-time dimensions but are required in dimensional regularization. We write the SM Lagrangian as
\begin{align}
	\label{eq:SM}
	\L _{\rm SM} &= -\frac14 G_{\mu \nu}^A G^{A\mu \nu}-\frac14 W_{\mu \nu}^I W^{I \mu \nu} -\frac14 B_{\mu \nu} B^{\mu \nu}
		+ (D_\mu H^\dagger)(D^\mu H)
		+\sum_{\mathclap{\psi=q,u,d,\ell,e}} \overline \psi\, i \slashed D \, \psi\nn
		&\quad -\lambda \left(H^\dagger H -\frac12 v^2\right)^2- \biggl[ H^{\dagger i} \overline d\, Y_d\, q_{i} + \widetilde H^{\dagger i} \overline u\, Y_u\, q_{i} + H^{\dagger i} \overline e\, Y_e\,  \ell_{i} + \hc \biggr] \nn
		&\quad + \L_\mathrm{GF} + \L_\mathrm{FP} \, ,
\end{align}
where $\widetilde H_i = \epsilon_{ij} H^{\dagger j}$ and the covariant derivative is defined as
\begin{equation}
	D_\mu = \p_\mu + i g_3 T^A G_\mu^A + i g_2 t^I W_\mu^I + i g_1 \y B_\mu \, ,
\end{equation}
with $SU(3)_c$ generators $T^A$, $SU(2)_L$ generators $t^I$, and $U(1)_Y$ hypercharge generator $\y$, see App.~\ref{sec:Conventions}. The Higgs doublet has hypercharge $\y_H = 1/2$. The Yukawa matrices $Y_{u,d,e}$ are $3\times3$ matrices in flavor space that can be expressed in terms of the CKM matrix and the mass matrices in the broken phase.
The renormalization can be performed in the unbroken phase and it is not necessary to consider spontaneous symmetry breaking (SSB)~\cite{Alonso:2013hga}. In particular, for the renormalization of dimension-six terms, all IR scales can be set to zero, including the vacuum expectation value $v$. At two loops, the Higgs self-interaction ($\lambda$) only leads to massless tadpoles, which are power-divergent scaleless integrals that vanish in dimensional regularization (they are also a special case of factorizable diagrams, which do not contribute to RGEs~\cite{Jenkins:2023rtg,Naterop:2025factorizable}). The $\lambda$-dependence of the RGE is expected to start at three loops.

In the BNV sector, bosons only appear as virtual particles; hence we use conventional $R_\xi$ gauge fixing without the introduction of background fields,
\begin{align}
	\L_\mathrm{GF} &= - \frac{1}{2\xi_3} (G^A)^2 - \frac{1}{2\xi_2} (W^I)^2 - \frac{1}{2\xi_1} (B)^2 \, , \nn
	G^A &= \p^\mu G_\mu^A \, , \quad
	W^I = \p^\mu W_\mu^I \, , \quad
	B = \p^\mu B_\mu \, ,
\end{align}
and the corresponding ghost Lagrangian reads
\begin{equation}
	\L_\mathrm{FP} = -\bar\eta_G^A \bigg[ \Box \delta^{AB} + g_3 \pl_\mu f^{ACB} G^{C\mu} \bigg] \eta_G^B - \bar\eta_W^I \bigg[ \Box \delta^{IJ} + g_2 \pl_\mu \epsilon^{IKJ} W^{K\mu} \bigg] \eta_W^J - \bar\eta_B \Box \eta_B \, .
\end{equation}

\begin{table}[t]

\begin{center}

\begin{minipage}[t]{5.2cm}
\renewcommand{\arraystretch}{1.5}
\begin{tabular}[t]{c|c}
\multicolumn{2}{c}{\boldmath{$\Delta B = \Delta L = 1$}} \\
\hline
$\Q_{duq\ell}$      & $\epsilon^{\alpha \beta \gamma} \epsilon^{ij} (d^T_{\alpha p} C u_{\beta r} ) (q^T_{\gamma i s} C \ell_{jt})  $  \\
$\Q_{qque}$      & $\epsilon^{\alpha \beta \gamma} \epsilon^{ij} (q^T_{\alpha i p} C q_{\beta j r} ) (u^T_{\gamma s} C e_{t})  $  \\
$\Q_{qqq\ell}$      & $\epsilon^{\alpha \beta \gamma} \epsilon^{il} \epsilon^{jk} (q^T_{\alpha i p} C q_{\beta j r} ) (q^T_{\gamma k s} C \ell_{l t})  $  \\
$\Q_{duue}$      & $\epsilon^{\alpha \beta \gamma} (d^T_{\alpha p} C u_{\beta r} ) (u^T_{\gamma s} C e_{t})  $  \\
\end{tabular}
\end{minipage}
\end{center}
\caption{The dimension-six baryon-number-violating operators in SMEFT. There are also the Hermitian conjugate operators with $\Delta B = \Delta L = -1$. The subscripts $p, r, s, t$ are flavor indices in a weak eigenbasis.}
\label{tab:Operators}
\end{table}

In the present work, we consider the BNV dimension-six operators $\Q_i$, listed in Table~\ref{tab:Operators}. Their flavor symmetries were discussed in detail in Ref.~\cite{Alonso:2014zka}. In particular, the symmetry
\begin{equation}
	\Q_{\substack{qque\\prst}} = \Q_{\substack{qque\\rpst}}
\end{equation}
holds in general, while in $D=4$ space-time dimensions, the relation~\cite{Abbott:1980zj,Alonso:2014zka}
\begin{equation}
	\label{eq:qqqlSymm}
	\Q_{\substack{qqq\ell\\prst}} + \Q_{\substack{qqq\ell\\rpst}} = \Q_{\substack{qqq\ell\\sprt}} + \Q_{\substack{qqq\ell\\srpt}}
\end{equation}
is valid. In Eq.~\eqref{eq:SMEFT}, we implicitly sum over the flavor indices and redundancies are avoided by requiring that the Wilson coefficients fulfill the flavor symmetries
\begin{align}
	\label{eq:qqqlSymmCoeffs}
	C_{\substack{qque\\prst}} &= C_{\substack{qque\\rpst}} \, , \nn
	C_{\substack{qqq\ell\\prst}} + C_{\substack{qqq\ell\\rpst}} &= C_{\substack{qqq\ell\\rspt}} + C_{\substack{qqq\ell\\srpt}} \, .
\end{align}
We follow Ref.~\cite{Alonso:2014zka} and perform a decomposition of the coefficients $C_{\substack{duue\\prst}}$ and $C_{\substack{qqq\ell\\prst}}$ into representations of the permutation group,\footnote{Note that the ordering of the indices in Eq.~\eqref{eq:qqqlSymmCoeffs} needs to be different from Eq.~\eqref{eq:qqqlSymm}. In Ref.~\cite{Alonso:2014zka}, the indices of the mixed-symmetry tensors $M_{\substack{qqq\ell\\prst}}$ and $N_{\substack{qqq\ell\\prst}}$ are not correct.}
\begin{align}
	\label{eq:CqqqlDecomposition}
	C_{\substack{duue\\prst}}^{(\pm)} &= \frac{1}{2} \left( C_{\substack{duue\\prst}} \pm C_{\substack{duue\\psrt}} \right) \, , \nn
	C_{\substack{qqq\ell\\prst}} &= S_{\substack{qqq\ell\\prst}} + A_{\substack{qqq\ell\\prst}} + M_{\substack{qqq\ell\\prst}} + N_{\substack{qqq\ell\\prst}} \, , \nn
	S_{\substack{qqq\ell\\prst}} &= \frac{1}{6} \left( C_{\substack{qqq\ell\\prst}} + C_{\substack{qqq\ell\\sprt}} + C_{\substack{qqq\ell\\rspt}} + C_{\substack{qqq\ell\\psrt}} +  C_{\substack{qqq\ell\\srpt}} + C_{\substack{qqq\ell\\rpst}} \right) \, , \nn
	A_{\substack{qqq\ell\\prst}} &= \frac{1}{6} \left( C_{\substack{qqq\ell\\prst}} + C_{\substack{qqq\ell\\sprt}} + C_{\substack{qqq\ell\\rspt}} - C_{\substack{qqq\ell\\psrt}} - C_{\substack{qqq\ell\\srpt}} - C_{\substack{qqq\ell\\rpst}} \right) \, , \nn
	M_{\substack{qqq\ell\\prst}} &= \frac{1}{3} \left( C_{\substack{qqq\ell\\prst}} - C_{\substack{qqq\ell\\sprt}} + C_{\substack{qqq\ell\\srpt}} - C_{\substack{qqq\ell\\rpst}} \right) \, , \nn
	N_{\substack{qqq\ell\\prst}} &= \frac{1}{3} \left( C_{\substack{qqq\ell\\prst}} - C_{\substack{qqq\ell\\rspt}} - C_{\substack{qqq\ell\\srpt}} + C_{\substack{qqq\ell\\rpst}} \right) \, ,
\end{align}
where $N_{\substack{qqq\ell\\prst}} = 0$ due to the flavor symmetry~\eqref{eq:qqqlSymmCoeffs}. Since the gauge interactions are flavor diagonal, their contributions to the RGEs respect the flavor symmetries and are diagonalized in flavor space in the decomposition~\eqref{eq:CqqqlDecomposition} at any loop order~\cite{Alonso:2014zka}.


\section{Scheme definition}
\label{sec:Scheme}

\subsection{Dimensional regularization}

As in the LEFT~\cite{Naterop:2025lzc,Aebischer:2025hsx}, the two-loop renormalization of the BNV sector of the SMEFT does not involve any ill-defined $\gamma_5$-odd traces. Therefore, we perform the calculation in naive dimensional regularization (NDR) in $D=4-2\varepsilon$ space-time dimensions with anti-commuting $\gamma_5$, avoiding the issue of spurious symmetry breaking of the 't~Hooft and Veltman scheme and the need for finite symmetry-restoring renormalizations~\cite{tHooft:1972tcz,Breitenlohner:1977hr}.

\subsection{Evanescent operators}

In the NDR scheme, evanescent operators in the BNV sector appear due to the fact that Chisholm and Fierz relations do not hold away from $D=4$ space-time dimensions. We employ the same generic evanescent scheme as Refs.~\cite{Dekens:2019ept,Naterop:2025lzc}. In particular, we define the Chisholm-evanescent operators
\begin{align}
	\E^{(2)}_{\substack{duq\ell \\ prst}} &= -\epsilon^{\alpha \beta \gamma} \epsilon^{ij} (d^T_{\alpha p} C \sigma^{\mu\nu} u_{\beta r} ) (q^T_{\gamma i s} C \sigma_{\mu\nu} \ell_{jt}) - 2 (1+2a_\mathrm{ev}) \varepsilon \, \Q_{\substack{duq\ell\\prst}} \, , \nn
	\color{lightgray} \E^{(2)}_{\substack{qque \\ prst}} &\color{lightgray}= -\epsilon^{\alpha \beta \gamma} \epsilon^{ij} (q^T_{\alpha i p} C \sigma^{\mu\nu} q_{\beta j r} ) (u^T_{\gamma s} C \sigma_{\mu\nu} e_{t}) - 2 (1+2a_\mathrm{ev}) \varepsilon \, \Q_{\substack{qque\\prst}} \, ,
\end{align}
as well as the Fierz-evanescent operators
\begin{align}
	\label{eq:FierzEvanOps}
	\E^{(F1)}_{\substack{qud\ell \\ srpt}} &= \epsilon^{\alpha \beta \gamma} \epsilon^{ij} (q^T_{\gamma i s} C \gamma^{\mu} u_{\beta r} ) (d^T_{\alpha p} C \gamma_{\mu} \ell_{jt} ) + 2 \, \Q_{\substack{duq\ell\\prst}} \, , \nn
	\E^{(F1)}_{\substack{qdu\ell \\ sprt}} &= \epsilon^{\alpha \beta \gamma} \epsilon^{ij} (q^T_{\gamma i s} C \gamma^{\mu} d_{\alpha p} ) (u^T_{\beta r} C \gamma_{\mu} \ell_{jt} )  + 2 \, \Q_{\substack{duq\ell\\prst}} \, , \nn
	\E^{(F1)}_{\substack{uqqe \\ srpt}} &= \epsilon^{\alpha \beta \gamma} \epsilon^{ij} (u^T_{\gamma s} C \gamma^{\mu} q_{\beta j r} ) (q^T_{\alpha i p} C \gamma_{\mu} e_{t} ) + 2 \, \Q_{\substack{qque\\prst}} \, , \nn
	\E^{(F)}_{\substack{qqq\ell \\ prst}} &= \Q_{\substack{qqq\ell \\ prst}} + \Q_{\substack{qqq\ell \\ rpst}} - \Q_{\substack{qqq\ell \\ sprt}} - \Q_{\substack{qqq\ell \\ srpt}} \, , \nn
	\E^{(F2)}_{\substack{qqq\ell \\ prst}} &= \epsilon^{\alpha \beta \gamma} \epsilon^{i\ell} \epsilon^{jk} (q^T_{\alpha i p} C \sigma^{\mu\nu} q_{\beta j r} ) (q^T_{\gamma k s} C \sigma_{\mu\nu}  \ell_{\ell t}) - 4 \, \Q_{\substack{qqq\ell \\ prst}} - 8 \, \Q_{\substack{qqq\ell \\ psrt}} \, , \nn
	\E^{(F2)}_{\substack{duue \\ prst}} &= \epsilon^{\alpha \beta \gamma} (d^T_{\alpha p} C \sigma^{\mu\nu} u_{\beta r} ) (u^T_{\gamma s} C \sigma_{\mu\nu}  e_{t}) - 4 \, \Q_{\substack{duue \\ prst}} + 8 \, \Q_{\substack{duue \\ psrt}} \, , \nn
	\color{lightgray}\E^{(F)}_{\substack{uude \\ srpt}} &\color{lightgray}= \epsilon^{\alpha \beta \gamma} (u^T_{\alpha s} C u_{\beta r} ) (d^T_{\gamma p} C e_{t}) - \Q_{\substack{duue\\prst}} + \Q_{\substack{duue \\psrt}} \, , \nn
	\color{lightgray}\E^{(F2)}_{\substack{uude \\ srpt}} &\color{lightgray}= \epsilon^{\alpha \beta \gamma} (u^T_{\alpha s} C \sigma^{\mu\nu} u_{\beta r} ) (d^T_{\gamma p} C \sigma_{\mu\nu}  e_{t}) - 4 \, \Q_{\substack{duue\\prst}} - 4 \, \Q_{\substack{duue \\psrt}} \, .
\end{align}
The subset of evanescent operators without tensor Dirac structures has been classified previously in Ref.~\cite{Fuentes-Martin:2022vvu} in the context of matching calculations. The operators shown in light gray are not generated at one loop from the insertion of physical operators. However, defining them is useful for the extraction of finite counterterms from insertions of evanescent operators into one-loop diagrams. At two loops, additional Chisholm-evanescent operators are generated containing three or four Dirac matrices in each bilinear (an odd number of Dirac matrices arises due to Yukawa interactions). They can be defined following Refs.~\cite{Dekens:2019ept,Naterop:2025lzc}. Both the finite counterterms that compensate evanescent insertions as well as divergent two-loop counterterms depend on the two-loop evanescent scheme choice. However, this scheme dependence drops out in the two-loop RGEs, which can only depend on the one-loop evanescent scheme.

As explained in Ref.~\cite{Naterop:2025lzc}, the two-loop calculation is simplified by setting $a_\mathrm{ev} = -1/2$. In this scheme, the coefficient of the Chisholm-evanescent operator $\E^{(2)}_{\substack{qque\\prst}}$ exhibits the flavor symmetry
\begin{equation}
	K^{(2)}_{\substack{qque\\prst}} = - K^{(2)}_{\substack{qque\\rpst}} \, ,
\end{equation}
which explains why $\E^{(2)}_{\substack{qque\\prst}}$ is not generated by gauge interactions at any loop order. We will reconstruct the scheme dependence parametrized by $a_\mathrm{ev}$ in a second step.

The coefficient of the Fierz-evanescent operator $\E^{(F)}_{qqq\ell}$ is constrained to have three types of symmetries, which correspond to the symmetric, antisymmetric, and mixed tensors in Eq.~\eqref{eq:CqqqlDecomposition}:
\begin{align}
	0 &= K^{(F)}_{\substack{qqq\ell\\prst}} + K^{(F)}_{\substack{qqq\ell\\sprt}} + K^{(F)}_{\substack{qqq\ell\\rspt}} + K^{(F)}_{\substack{qqq\ell\\psrt}} + K^{(F)}_{\substack{qqq\ell\\srpt}} + K^{(F)}_{\substack{qqq\ell\\rpst}} \, , \nn
	0 &= K^{(F)}_{\substack{qqq\ell\\prst}} + K^{(F)}_{\substack{qqq\ell\\sprt}} + K^{(F)}_{\substack{qqq\ell\\rspt}} - K^{(F)}_{\substack{qqq\ell\\psrt}} - K^{(F)}_{\substack{qqq\ell\\srpt}} - K^{(F)}_{\substack{qqq\ell\\rpst}} \, , \nn
	0 &= K^{(F)}_{\substack{qqq\ell\\prst}} - K^{(F)}_{\substack{qqq\ell\\sprt}} + K^{(F)}_{\substack{qqq\ell\\srpt}} - K^{(F)}_{\substack{qqq\ell\\rpst}} \, .
\end{align}
Finally, the coefficients of the last two Fierz-evanescent operators in Eq.~\eqref{eq:FierzEvanOps} exhibit the flavor symmetries
\begin{equation}
	K^{(F)}_{\substack{uude\\srpt}} = -K^{(F)}_{\substack{uude\\rspt}} \, , \quad K^{(F2)}_{\substack{uude\\srpt}} = K^{(F2)}_{\substack{uude\\rspt}} \, .
\end{equation}


\section{Renormalization}
\label{sec:Renormalization}

\subsection{One-loop renormalization}

Following the conventions of Ref.~\cite{Naterop:2024cfx}, we write the RGEs in the form
\begin{equation}
	\label{eq:RGENotation}
	\dot X = \frac{d}{d\log\mu} X = \frac{1}{16\pi^2} [\dot X]_1 + \frac{1}{(16\pi^2)^2} [\dot X]_2 \, .
\end{equation}
The one-loop RGEs of the BNV dimension-six operators were obtained in Refs.~\cite{Abbott:1980zj,Alonso:2014zka} and we agree with these results.\footnote{In the case of the Yukawa contributions to the RGE of $C_{qqq\ell}$, Ref.~\cite{Alonso:2014zka} gives the result without splitting off the evanescent component $N_{qqq\ell}$, i.e., their result does not manifestly fulfill the symmetry~\eqref{eq:qqqlSymmCoeffs}.} In terms of the representations of the flavor-permutation group~\eqref{eq:CqqqlDecomposition}, they read
\begin{align}
	\left[ \dot C_{\substack{duq\ell\\prst}} \right]_1 &= \left( 6(\y_d\y_u + \y_\ell\y_q) g_1^2 - \frac{9}{2} g_2^2 - 4 g_3^2 \right) C_{\substack{duq\ell\\prst}}  - (Y_u^\dagger)_{wr} (Y_u)_{vs} C_{\substack{duq\ell\\pvwt}} - (Y_d^\dagger)_{wp} (Y_d)_{vs} C_{\substack{duq\ell\\vrwt}} \nn
		&\quad  + ( Y_d Y_d^\dagger )_{vp} C_{\substack{duq\ell\\vrst}}  + ( Y_u Y_u^\dagger )_{vr} C_{\substack{duq\ell\\pvst}} + \frac{1}{2} ( Y_e^\dagger Y_e)_{vt} C_{\substack{duq\ell\\prsv}} + \frac{1}{2} ( Y_u^\dagger Y_u + Y_d^\dagger Y_d )_{vs} C_{\substack{duq\ell\\prvt}} \nn
		&\quad + 2 (Y_d^\dagger)_{vp} (Y_e)_{wt} C_{\substack{qque\\vsrw}} + 3 (Y_u)_{vs} (Y_e)_{wt} C_{\substack{duue\\prvw}}^{(+)} +  (Y_u)_{vs} (Y_e)_{wt} C_{\substack{duue\\prvw}}^{(-)} \nn
		&\quad + (Y_d^\dagger)_{vp} (Y_u^\dagger)_{wr} \Big( 6 S_{\substack{qqq\ell\\vwst}} + 3 M_{\substack{qqq\ell\\vwst}} - 6 M_{\substack{qqq\ell\\vswt}} \Big)  \, , \nn
	\left[ \dot C_{\substack{qque\\prst}} \right]_1 &= \left( 6(\y_e\y_u + \y_q^2) g_1^2 - \frac{9}{2} g_2^2 - 4 g_3^2 \right) C_{\substack{qque\\prst}} + \frac{1}{2} (Y_d)_{vp} (Y_e^\dagger)_{wt} C_{\substack{duq\ell\\vsrw}} + \frac{1}{2} (Y_d)_{vr} (Y_e^\dagger)_{wt} C_{\substack{duq\ell\\vspw}} \nn
		&\quad + (Y_e Y_e^\dagger)_{vt} C_{\substack{qque\\prsv}} + (Y_u Y_u^\dagger)_{vs} C_{\substack{qque\\prvt}} + \frac{1}{2} (Y_u^\dagger Y_u + Y_d^\dagger Y_d)_{vp} C_{\substack{qque\\vrst}} \nn
		&\quad + \frac{1}{2} (Y_u^\dagger Y_u + Y_d^\dagger Y_d)_{vr} C_{\substack{qque\\pvst}} - (Y_u)_{vp} (Y_u^\dagger)_{ws} C_{\substack{qque\\rwvt}} - (Y_u)_{vr} (Y_u^\dagger)_{ws} C_{\substack{qque\\pwvt}} \nn
		&\quad - \frac{1}{2} \Big( (Y_d)_{vr} (Y_u)_{wp} + (Y_d)_{vp} (Y_u)_{wr} \Big) \Big( 3 C_{\substack{duue\\vwst}}^{(+)} + C_{\substack{duue\\vwst}}^{(-)} \Big) \nn
		&\quad - (Y_u^\dagger)_{vs} (Y_e^\dagger)_{wt} \Big( 3 S_{\substack{qqq\ell\\prvw}} + \frac{3}{2} M_{\substack{qqq\ell\\prvw}} - 3 M_{\substack{qqq\ell\\pvrw}} \Big)  \, , \nn
	\left[ \dot C_{\substack{duue\\prst}}^{(+)} \right]_1 &= \left( -2(\y_d\y_e + \y_u^2 - 4 \y_u\y_d - 4\y_u\y_e) g_1^2 - 4 g_3^2 \right) C_{\substack{duue\\prst}}^{(+)} \nn
		&\quad + 2 (Y_u^\dagger)_{vr} (Y_e^\dagger)_{wt} C_{\substack{duq\ell\\psvw}} + 2 (Y_u^\dagger)_{vs} (Y_e^\dagger)_{wt} C_{\substack{duq\ell\\prvw}} \nn
		&\quad - 4 (Y_d^\dagger)_{vp} (Y_u^\dagger)_{wr} C_{\substack{qque\\vwst}} - 4 (Y_d^\dagger)_{vp} (Y_u^\dagger)_{ws} C_{\substack{qque\\vwrt}} \nn
		&\quad + (Y_d Y_d^\dagger)_{vp} C_{\substack{duue\\vrst}}^{(+)} + (Y_u Y_u^\dagger)_{vr} C_{\substack{duue\\pvst}}^{(+)} + (Y_u Y_u^\dagger)_{vs} C_{\substack{duue\\prvt}}^{(+)} + (Y_e Y_e^\dagger)_{vt} C_{\substack{duue\\prsv}}^{(+)} \, , \nn
	\left[ \dot C_{\substack{duue\\prst}}^{(-)} \right]_1 &= \left( 6(\y_d\y_e + \y_u^2) g_1^2 - 4 g_3^2 \right) C_{\substack{duue\\prst}}^{(-)} \nn*
		&\quad - 2 (Y_u^\dagger)_{vr} (Y_e^\dagger)_{wt} C_{\substack{duq\ell\\psvw}} + 2 (Y_u^\dagger)_{vs} (Y_e^\dagger)_{wt} C_{\substack{duq\ell\\prvw}} \nn
		&\quad - 4 (Y_d^\dagger)_{vp} (Y_u^\dagger)_{wr} C_{\substack{qque\\vwst}} + 4 (Y_d^\dagger)_{vp} (Y_u^\dagger)_{ws} C_{\substack{qque\\vwrt}} \nn
		&\quad + (Y_d Y_d^\dagger)_{vp} C_{\substack{duue\\vrst}}^{(-)} + (Y_u Y_u^\dagger)_{vr} C_{\substack{duue\\pvst}}^{(-)} + (Y_u Y_u^\dagger)_{vs} C_{\substack{duue\\prvt}}^{(-)} + (Y_e Y_e^\dagger)_{vt} C_{\substack{duue\\prsv}}^{(-)} \, , \nn
	\left[ \dot S_{\substack{qqq\ell\\prst}} \right]_1 &= \left( 6 \y_q(\y_q + \y_\ell) g_1^2 - 15 g_2^2 - 4 g_3^2 \right) S_{\substack{qqq\ell\\prst}} + \frac{1}{2} (Y_d^\dagger Y_d + Y_u^\dagger Y_u)_{vp}  S_{\substack{qqq\ell\\vrst}} \nn
		&\quad + \frac{1}{2} (Y_d^\dagger Y_d + Y_u^\dagger Y_u)_{vr}  S_{\substack{qqq\ell\\pvst}} + \frac{1}{2} (Y_d^\dagger Y_d + Y_u^\dagger Y_u)_{vs}  S_{\substack{qqq\ell\\prvt}} + \frac{1}{2} (Y_e^\dagger Y_e)_{vt}  S_{\substack{qqq\ell\\prsv}} \nn
		&\quad + \frac{2}{3} \Big( (Y_u)_{wp} (Y_d)_{vr} + (Y_u)_{wr} (Y_d)_{vp} \Big) C_{\substack{duq\ell\\vwst}} + \frac{2}{3} \Big( (Y_u)_{wr} (Y_d)_{vs} + (Y_u)_{ws} (Y_d)_{vr} \Big) C_{\substack{duq\ell\\vwpt}} \nn
		&\quad + \frac{2}{3} \Big( (Y_u)_{wp} (Y_d)_{vs} + (Y_u)_{ws} (Y_d)_{vp} \Big) C_{\substack{duq\ell\\vwrt}} - \frac{4}{3} (Y_u)_{vp} (Y_e)_{wt} C_{\substack{qque\\rsvw}} \nn
 		&\quad - \frac{4}{3} (Y_u)_{vr} (Y_e)_{wt} C_{\substack{qque\\psvw}} - \frac{4}{3} (Y_u)_{vs} (Y_e)_{wt} C_{\substack{qque\\prvw}} \, , \nn
	\left[ \dot A_{\substack{qqq\ell\\prst}} \right]_1 &= \left( 6 \y_q(\y_q + \y_\ell) g_1^2 + 9 g_2^2 - 4 g_3^2 \right) A_{\substack{qqq\ell\\prst}} + \frac{1}{2} (Y_d^\dagger Y_d + Y_u^\dagger Y_u)_{vp}  A_{\substack{qqq\ell\\vrst}} \nn
		&\quad + \frac{1}{2} (Y_d^\dagger Y_d + Y_u^\dagger Y_u)_{vr}  A_{\substack{qqq\ell\\pvst}} + \frac{1}{2} (Y_d^\dagger Y_d + Y_u^\dagger Y_u)_{vs}  A_{\substack{qqq\ell\\prvt}} + \frac{1}{2} (Y_e^\dagger Y_e)_{vt}  A_{\substack{qqq\ell\\prsv}} \, , \nn
	\left[ \dot M_{\substack{qqq\ell\\prst}} \right]_1 &= \left( 6 \y_q(\y_q + \y_\ell) g_1^2 - 3 g_2^2 - 4 g_3^2 \right) M_{\substack{qqq\ell\\prst}} + \frac{1}{2} (Y_d^\dagger Y_d + Y_u^\dagger Y_u)_{vp}  M_{\substack{qqq\ell\\vrst}} \nn
		&\quad + \frac{1}{2} (Y_d^\dagger Y_d + Y_u^\dagger Y_u)_{vr}  M_{\substack{qqq\ell\\pvst}} + \frac{1}{2} (Y_d^\dagger Y_d + Y_u^\dagger Y_u)_{vs}  M_{\substack{qqq\ell\\prvt}} + \frac{1}{2} (Y_e^\dagger Y_e)_{vt}  M_{\substack{qqq\ell\\prsv}} \nn
		&\quad + \frac{2}{3} \Big( (Y_u)_{wr} (Y_d)_{vs} + (Y_u)_{ws} (Y_d)_{vr} \Big) C_{\substack{duq\ell\\vwpt}} - \frac{2}{3} \Big( (Y_u)_{wp} (Y_d)_{vs} + (Y_u)_{ws} (Y_d)_{vp} \Big) C_{\substack{duq\ell\\vwrt}} \nn
		&\quad - \frac{4}{3} (Y_u)_{vp} (Y_e)_{wt} C_{\substack{qque\\rsvw}} + \frac{4}{3} (Y_u)_{vr} (Y_e)_{wt} C_{\substack{qque\\psvw}} \, .
\end{align}
Note that even including the Yukawa contributions, the coefficient $A_{qqq\ell}$ decouples at one loop from all the other Wilson coefficients.

In the NDR scheme, one-loop insertions of physical operators (we use the abbreviation $\{X\}_1 = X/(16\pi^2)$) generate the evanescent divergences
\begin{align}
	\delta_\mathrm{div} \bigg( K^{(2)}_{\substack{duq\ell\\prst}} \bigg) &= \left\{ \frac{g_1^2 (\y_d - \y_u)(\y_q - \y_\ell)}{4\varepsilon} \right\}_1 C_{\substack{duq\ell\\prst}} \, , \nn
	\delta_\mathrm{div} \bigg( K^{(F1)}_{\substack{qud\ell\\srpt}} \bigg) &= \left\{ \frac{1}{4\varepsilon} \right\}_1 \left( C_{\substack{duq\ell\\vrwt}} (Y_d)_{vs} (Y_d^\dagger)_{wp} - \Bigl( 3 A_{\substack{qqq\ell\\svwt}} + 3 M_{\substack{qqq\ell\\svwt}} + S_{\substack{qqq\ell\\svwt}} \Bigr) (Y_u^\dagger)_{vr} (Y_d^\dagger)_{wp} \right)  \, , \nn
	\delta_\mathrm{div} \bigg( K^{(F1)}_{\substack{qdu\ell\\sprt}} \bigg) &= \left\{ \frac{1}{4\varepsilon} \right\}_1 \begin{aligned}[t]
		& \biggl( C_{\substack{duq\ell\\pvwt}} (Y_u)_{vs} (Y_u^\dagger)_{wr} - \left( C_{\substack{duue\\pvrw}}^{(+)} + C_{\substack{duue\\pvrw}}^{(-)} \right) (Y_u)_{vs} (Y_e)_{wt} \\
		&\quad - 2 C_{\substack{qque\\vsrw}} (Y_d^\dagger)_{vp} (Y_e)_{wt}  - \left( 3 A_{\substack{qqq\ell\\svwt}} + 3 M_{\substack{qqq\ell\\svwt}} +  S_{\substack{qqq\ell\\svwt}} \right) (Y_d^\dagger)_{vp} (Y_u^\dagger)_{wr} \biggr)  \, , \end{aligned} \nn
	\delta_\mathrm{div} \bigg( K^{(F1)}_{\substack{uqqe\\srpt}} \bigg) &= \left\{ \frac{1}{4\varepsilon} \right\}_1 \begin{aligned}[t]
		& \biggl( 2 C_{\substack{qque\\vrwt}} (Y_u^\dagger)_{vs} (Y_u)_{wp} + \left( C_{\substack{duue\\vswt}}^{(+)} + C_{\substack{duue\\vswt}}^{(-)} \right) (Y_d)_{vr} (Y_u)_{wp}  \\
		&\quad - C_{\substack{duq\ell\\vspw}} (Y_d)_{vr} (Y_e^\dagger)_{wt} + ( 3 A_{\substack{qqq\ell\\vrpw}} + 3 
    M_{\substack{qqq\ell\\vrpw}} +  
    S_{\substack{qqq\ell\\vrpw}} ) (Y_u^\dagger)_{vs} (Y_e^\dagger)_{wt}
	\biggr)  \, , \end{aligned} 
    \nn
	\delta_\mathrm{div} \bigg( K^{(F)}_{\substack{qqq\ell\\prst}} \bigg) &= \left\{ \frac{1}{4\varepsilon} \right\}_1 \begin{aligned}[t]
		& \frac{8}{9} \biggl(  C_{\substack{duq \ell\\ w v s t}} (Y_u)_{v p} (Y_d)_{w r}  -  C_{\substack{duq \ell\\ w v p t}} (Y_u)_{v r} (Y_d)_{w s} \\
		&\quad + C_{\substack{duq \ell\\ v w s t}} (Y_u)_{w r} (Y_d)_{v p} -  C_{\substack{duq \ell\\ v w p t}} (Y_u)_{w s} (Y_d)_{v r} \\
	        &\quad + 2 C_{\substack{qque\\ r s v w}} (Y_u)_{v p} (Y_e)_{w t} - 2 C_{\substack{qque\\ p r  v w}} (Y_u)_{v s} (Y_e)_{w t} - 9 g^2_2  M_{\substack{qqq\ell\\ p s r t}}
	\biggr)  \, ,  \end{aligned}  
     \nn
	\delta_\mathrm{div} \bigg( K^{(F2)}_{\substack{qqq\ell\\prst}} \bigg) &= - \left\{ \frac{g_2^2}{4\varepsilon} \right\}_1 \left( 3 A_{\substack{qqq\ell\\prst}} + 3 
    M_{\substack{qqq\ell\\prst}} +  
    S_{\substack{qqq\ell\\prst}} \right) \, , \nn
	\delta_\mathrm{div} \bigg( K^{(F2)}_{\substack{duue\\prst}} \bigg) &= \left\{ \frac{g_1^2 (\y_d - \y_u)(\y_e - \y_u)}{4\varepsilon} \right\}_1 \left( C_{\substack{duue\\prst}}^{(+)} + C_{\substack{duue\\prst}}^{(-)} \right) \, ,
\end{align}
where the $U(1)_Y$ contributions agree with the results of Ref.~\cite{Naterop:2025lzc} for $U(1)_{\rm em}$.

\subsection{Two-loop renormalization}

To obtain two-loop RGEs that avoid a mixing of the evanescent operators into the physical sector, the evanescent operators that are generated at one loop need to be renormalized~\cite{Dugan:1990df,Herrlich:1994kh}. We insert them again into one-loop diagrams to determine the finite physical counterterms that compensate their one-loop effect~\cite{Fuentes-Martin:2022vvu}.
These finite renormalizations directly enter the RGEs for the physical Wilson coefficients $C_i$, which are obtained from the master formula~\cite{Naterop:2023dek}
\begin{equation}
	\label{eq:RGEMasterFormula}
	\left[ \, \dot C_i \, \right]_2 = 
		4 C_i^{(2,1)} - 2 \sum_j  K_j^{(1,1)} \frac{\p C_i^{(1,0)}}{\p K_j} \, ,
\end{equation}
with $X_i^{(l,n)}$, $X\in\{C,K\}$, denoting the coefficients of the $1/\varepsilon^n$ poles of the $l$-loop counterterms defined by
\begin{equation}
	X_i^\mathrm{ct} = \sum_{l=1}^\infty \sum_{n=0}^l \frac{1}{\varepsilon^n} \frac{1}{(16\pi^2)^l} X_i^{(l,n)}(\{L_j^r(\mu)\},\{K_k^r(\mu)\})
\end{equation}
and the sum in Eq.~\eqref{eq:RGEMasterFormula} runs over the coefficients of all evanescent operators. The RGEs are determined from the $1/\varepsilon$ poles of the two-loop counterterms, apart from the correction term due to the finite renormalizations.

Our two-loop calculation relies on two independent implementations. The first one consists of a tool chain adapted from Refs.~\cite{Naterop:2023dek,Naterop:2024cfx,Naterop:2025lzc,Naterop:2025cwg}: diagrams are generated with \texttt{QGRAF}~\cite{Nogueira:1991ex}. The application of Feynman rules, $SU(n)$, Dirac, and Lorentz algebra are performed with our own routines written in \texttt{Mathematica} and \texttt{FORM}~\cite{Vermaseren:2000nd,Ruijl:2017dtg}. The two-loop integrals are treated with the infrared rearrangement in terms of the local $R$-operation in combination with the introduction of an auxiliary mass~\cite{Chetyrkin:1997fm}. The second implementation relies on \texttt{FeynRules}~\cite{Alloul:2013bka}, \texttt{FeynArts}~\cite{Hahn:2000kx}, \texttt{FeynCalc}~\cite{Mertig:1990an,Shtabovenko:2016sxi}, as well as \texttt{FIRE}~\cite{Smirnov:2023yhb} for the reduction of two-loop integrals into a small set of master integrals, which are evaluated with Mellin--Barnes techniques~\cite{Czakon:2005rk,Smirnov:2009up}.

\subsection{Scheme dependence}

Although the natural evanescent scheme choice in the BNV sector is $a_\mathrm{ev} = -1/2$ which simplifies the two-loop calculation, the dependence on a generic scheme parameter can be easily recovered a posteriori through a basis change~\cite{Naterop:2025lzc}. Instead of doing the calculation with an evanescent operator with generic $a_\mathrm{ev}$
\begin{equation}
	\E^{(2),a_\mathrm{ev}}_{\substack{duq\ell \\ prst}} = \E^{(2)}_{\substack{duq\ell \\ prst}} - 2 (1+2a_\mathrm{ev}) \varepsilon \, \Q_{\substack{duq\ell\\prst}} \, , \nn
\end{equation}
we use the operator basis with $a_\mathrm{ev} = -1/2$ and perform a shift in the Wilson coefficient
\begin{equation}
	C_{\substack{duq\ell \\ prst}} = C^{a_\mathrm{ev}}_{\substack{duq\ell \\ prst}} - 2 (1+2a_\mathrm{ev}) \varepsilon \, K^{(2),a_\mathrm{ev}}_{\substack{duq\ell\\prst}} \, , \quad K^{(2)}_{\substack{duq\ell\\prst}} = K^{(2),a_\mathrm{ev}}_{\substack{duq\ell\\prst}} \, ,
\end{equation}
which corresponds to a finite renormalization~\cite{Naterop:2025lzc}
\begin{equation}
	\delta_\mathrm{fin}^{a_\mathrm{ev}} \Bigl( C_{\substack{duq\ell \\ prst}} \Bigr) = \left\{ (1+2a_\mathrm{ev}) \frac{g_1^2 (\y_d - \y_u)(\y_\ell-\y_q)}{2} \right\}_1 C_{\substack{duq\ell \\ prst}} \, .
\end{equation}
This affects the two-loop RGEs in multiple ways: besides the insertion into one-loop counterterm diagrams, which changes the two-loop counterterm $C_i^{(2,1)}$, there are two additional correction terms to the RGE master formula~\cite{Naterop:2023dek}
\begin{equation}
	\label{eq:RGEMasterFormulaFull}
	\left[ \, \dot C_i \, \right]_2 = 
		4 C_i^{(2,1)} - 2 \sum_j  C_j^{(1,1)} \frac{\p C_i^{(1,0)}}{\p C_j} - 2 \sum_j  C_j^{(1,0)} \frac{\p C_i^{(1,1)}}{\p C_j} - 2 \sum_j  K_j^{(1,1)} \frac{\p C_i^{(1,0)}}{\p K_j} \, ,
\end{equation}
where the sum over $C_j$ in the second term runs over physical parameters, both Wilson coefficients and gauge couplings. The pure gauge contribution to the $a_\mathrm{ev}$-dependence can be derived from Ref.~\cite{Naterop:2025lzc},
\begin{equation}
	\left[ \dot C_{\substack{duq\ell\\prst}} \right]_2^{a_\mathrm{ev}} = (1+2a_\mathrm{ev}) b_0^{g_1} g_1^4 (\y_d - \y_u)(\y_\ell - \y_q) C_{\substack{duq\ell\\prst}} \, ,
\end{equation}
where $b_0^{g_1}$ is the coefficient of the one-loop $U(1)_Y$ beta function, defined below. In addition to pure $U(1)_Y$ gauge contributions, the scheme dependence of the RGEs, parametrized by $a_\mathrm{ev}$, contains mixed $U(1)_Y$--Yukawa contributions. We include the full $a_\mathrm{ev}$-dependence in our final results, which are provided as supplementary material in electronic form.


\section{Results for the two-loop RGEs}
\label{sec:Results}

The dependence on the number of fermion flavors $n_f$ arises from vacuum-polarization diagrams and is conveniently written in terms of the coefficients of the one-loop beta functions
\begin{align}
	b_0^{g_1} &= - \frac{2}{3} \left( n_e \y_e^2 + 2 n_\ell \y_\ell^2 + N_c \left( n_u \y_u^2 + n_d \y_d^2 + 2 n_q \y_q^2 \right) + \y_H^2 \right)  \, , \nn
	b_0^{g_2} &= \frac{22}{3} - \frac{1}{3} \left( n_\ell + N_c n_q + \frac{1}{2} \right) \, , \nn
	b_0^{g_3} &= \frac{11}{3} N_c - \frac{1}{3} \left( n_u + n_d + 2 n_q \right) \, ,
\end{align}
with $N_c = 3$. In the absence of Yukawa interactions, the results are much simpler, in particular there is no mixing between operators with different chiralities. The RGEs for the purely right-chiral and the purely left-chiral operator coefficients $C_{duue}$ and $C_{qqq\ell}$ are most conveniently decomposed into the different representations of flavor permutations~\eqref{eq:CqqqlDecomposition}, which diagonalizes the gauge contributions. Evaluated for physical hypercharges and the case of three fermion generations, we find
\begin{align}
	\left[ \dot C_{\substack{duq\ell\\prst}} \right]_2 &= \left( \frac{1}{144} (723-1312 a_\mathrm{ev}) g_1^4+\frac{3 g_1^2 g_2^2}{8}+\frac{25 g_1^2 g_3^2}{9}-\frac{541 g_2^4}{16}+9 g_2^2 g_3^2-\frac{62 g_3^4}{3} \right) C_{\substack{duq\ell\\prst}} + \ldots \, , \nn
	\left[ \dot C_{\substack{qque\\prst}} \right]_2 &= \left( \frac{2803 g_1^4}{144} + \frac{7 g_1^2 g_2^2}{8} - \frac{541 g_2^4}{16} + \frac{70 g_1^2 g_3^2}{9} + 6 g_2^2 g_3^2 - \frac{62 g_3^4}{3} \right) C_{\substack{qque\\prst}} + \ldots \, , \nn
	\left[ \dot C_{\substack{duue\\prst}}^{(+)} \right]_2 &=  \left( \frac{265 g_1^4}{9} + \frac{68 g_1^2 g_3^2}{3} - \frac{22 g_3^4}{3} \right) C_{\substack{duue\\prst}}^{(+)} + \ldots \, , \nn
	\left[ \dot C_{\substack{duue\\prst}}^{(-)} \right]_2 &=  \left( -\frac{395 g_1^4}{27} - \frac{28 g_1^2 g_3^2}{3} - \frac{22 g_3^4}{3} \right) C_{\substack{duue\\prst}}^{(-)} + \ldots \, , \nn
	\left[ \dot S_{\substack{qqq\ell\\prst}} \right]_2 &= \left( - \frac{115 g_1^4}{216} + \frac{31 g_1^2 g_2^2}{12} - \frac{3427 g_2^4}{24} + \frac{7 g_1^2 g_3^2}{3} + 31 g_2^2 g_3^2 - \frac{22 g_3^4}{3} \right) S_{\substack{qqq\ell\\prst}} + \ldots \, , \nn
	\left[ \dot A_{\substack{qqq\ell\\prst}} \right]_2 &= \left( -\frac{115 g_1^4}{216} - \frac{11 g_1^2 g_2^2}{4} + \frac{533 g_2^4}{24} + \frac{7 g_1^2 g_3^2}{3} - 33 g_2^2 g_3^2 - \frac{22 g_3^4}{3} \right) A_{\substack{qqq\ell\\prst}} + \ldots  \, , \nn
	\left[ \dot M_{\substack{qqq\ell\\prst}} \right]_2 &= \left(-\frac{115 g_1^4}{216} - \frac{g_1^2 g_2^2}{12} + \frac{271 g_2^4}{8} + \frac{7 g_1^2 g_3^2}{3} - g_2^2 g_3^2 - \frac{22 g_3^4}{3} \right) M_{\substack{qqq\ell\\prst}} + \ldots \, .
\end{align}
where the ellipses stand for Yukawa and mixed Yukawa--gauge contributions. The complete results for the two-loop RGEs for the BNV dimension-six operators are provided in electronic form as supplementary material. There, we include the full flavor structure due to the Yukawa contributions and the explicit dependence on the hypercharges $\y_f$.

In order to ensure the validity of the results, we have performed several checks and compared to partial results existing in the literature. First, the entire calculation is performed in a generic $R_\xi$ gauge. The cancellation of the gauge-parameter dependence provides an important check of the calculation. As usual in the BNV sector, the gauge-parameter cancellation only happens if one sets $N_c=3$ as well as hypercharge conservation. Second, we have used generic parameters for the two-loop evanescent scheme. The two-loop RGEs are allowed to depend only on evanescent-scheme parameters that arise at one loop, which in the present case is only $a_\mathrm{ev}$. The cancellation of the dependence on two-loop evanescent-scheme parameters between the two terms in Eq.~\eqref{eq:RGEMasterFormula} provides another check. Third, we have compared the one-loop RGEs to Ref.~\cite{Alonso:2014zka}, as discussed above. Finally, we have compared and validated the $SU(3)_c \times U(1)_Y$ two-loop contributions to the RGEs with the analogous calculation in the LEFT~\cite{Naterop:2025lzc}, finding full agreement. This check is facilitated by keeping generic hypercharges, which can then be adjusted to match the $U(1)_\mathrm{em}$ charges in the LEFT.


\section{\boldmath Matching of the $S_1$ leptoquark model onto the SMEFT}
\label{sec:Matching}

In the following, we match the model with the scalar $SU(2)_L$ singlet leptoquark with hypercharge $\y_{S_1} = -1/3$ ($S_1$) onto the SMEFT.\footnote{See Ref.~\cite{Aebischer:2018acj} for QCD corrections to a similar LQ matching and Ref.~\cite{Gherardi:2020det} for the $S_1+S_3$ LQ matching without BNV.} Interestingly, the $S_1$ model can generate all Wilson coefficients of the four dimension-six BNV operators of the SMEFT. This will thus allow us to illustrate the cancellation of the matching-scale and of scheme dependences between one-loop matching contributions and the two-loop RGE.

\subsection{Tree-level matching}

In addition to gauge-kinetic terms and the mass $M$, we only include the fermion interactions of $S_1$, while neglecting Higgs- and LQ self-interactions.\footnote{LQ self and Higgs interactions are free parameters and not needed for checking the scale cancellation between the matching and the RGE.} Slightly simplifying the conventions of Ref.~\cite{Crivellin:2021ejk}, we have the Lagrangian
\begin{align}
	\label{eq:LeptoQuarkLagrangian}
	\L_{S_1}^Y &=  Y_{rp}^{RR} \; S_{1}^{\dagger}  (u_p^T C e_r)
				+ Y_{pr}^{LL} \; S_{1}^{\dagger} \epsilon^{ij} (q_{ip}^T C \ell_{jr}) \nn
				&\quad + Y_{pr}^{Q, LL \dagger} \; \epsilon^{\alpha\beta\gamma} \epsilon^{ij} S_{1, \alpha} ( q_{\beta i p}^T C q_{\gamma j r} )
				+ Y_{rp}^{Q, RR \dagger} \; \epsilon^{\alpha\beta\gamma} S_{1, \alpha} (d_{\beta p}^T C u_{\gamma r} ) + \hc \, .
\end{align}
Here, the Yukawa-like couplings $Y$ are arbitrary matrices, except for $Y_{pr}^{Q,LL}$, which can be chosen to be symmetric in flavor space without loss of generality ($Y_{pr}^{Q,LL} = Y_{rp}^{Q,LL}$), and we use the notation $Y^\dagger_{pr} = Y^*_{rp}$.
For the tree-level matching onto SMEFT at dimension six, one obtains for the coefficients of the physical operators~\cite{Dorsner:2016wpm,deBlas:2017xtg}\footnote{Ref.~\cite{deBlas:2017xtg} gives the matching result for $C_{qqq\ell}$ without splitting off the evanescent component $N_{qqq\ell}$, i.e., their result does not fulfill the flavor symmetry~\eqref{eq:qqqlSymmCoeffs}.}
\begin{align}
	\label{eq:TreeLevelMatching}
	C_{\substack{duq\ell\\prst}} &=  \frac{Y^{Q,RR \dagger}_{rp} Y^{LL}_{st}}{M^2} \, , \qquad
	C_{\substack{qque\\prst}} = \frac{Y^{Q,LL \dagger}_{pr} Y^{RR}_{ts}}{M^2} \, , \qquad
	C_{\substack{duue\\prst}} = \frac{Y^{Q,RR \dagger}_{rp} Y^{RR}_{ts}}{M^2} \, , \nn
	C_{\substack{qqq\ell\\prst}} &= -\frac{2 \, \left( Y^{Q,LL \dagger}_{pr} Y^{LL}_{st} + 2 Y^{Q,LL \dagger}_{rs} Y^{LL}_{pt} \right)}{3 M^2} \, ,
\end{align}
with the decomposition into the representations of the permutation group in Eq.~\eqref{eq:CqqqlDecomposition}
\begin{align}
	\label{eq:TreeLevelMatching2}
	C_{\substack{duue\\prst}}^{(\pm)} &= \frac{Y^{Q,RR \dagger}_{rp} Y^{RR}_{ts} \pm Y^{Q,RR \dagger}_{sp} Y^{RR}_{tr}}{2M^2} \, , \nn
	S_{\substack{qqq\ell\\prst}} &= -\frac{2( Y^{Q,LL \dagger}_{pr} Y^{LL}_{st} + Y^{Q,LL \dagger}_{ps} Y^{LL}_{rt} + Y^{Q,LL \dagger}_{rs} Y^{LL}_{pt})}{3M^2} \, , \quad A_{\substack{qqq\ell\\prst}} = 0 \, , \nn
	M_{\substack{qqq\ell\\prst}} &= \frac{2( Y^{Q,LL \dagger}_{ps} Y^{LL}_{rt} - Y^{Q,LL \dagger}_{rs} Y^{LL}_{pt})}{3M^2} \, , \quad N_{\substack{qqq\ell\\prst}} = 0 \, .
\end{align}
Doing the matching calculation at $D\neq4$ space-time dimensions shows that in the purely left-chiral sector a matching contribution to the Fierz-evanescent operator $\E^{(F)}_{qqq\ell}$ is obtained already at tree level,
\begin{equation}
	K^{(F)}_{\substack{qqq\ell\\prst}} = \frac{4 \, \left( Y^{Q,LL \dagger}_{rs} Y^{LL}_{pt}  - Y^{Q,LL \dagger}_{pr} Y^{LL}_{st} \right)}{9 M^2} \, ,
\end{equation}
which produces a physical effect in the one-loop matching.

\subsection{One-loop matching}

For the one-loop matching, we renormalize the parameters of the LQ Lagrangian in the \msbar{} scheme. The SMEFT is renormalized in the \msbar{} scheme as well, with the exception of the finite counterterms that compensate the physical effects of evanescent insertions~\cite{Dugan:1990df,Herrlich:1994kh}.\footnote{This is equivalent to performing the calculation in the EFT without inserting evanescent operators into loop diagrams and neglecting the finite counterterms, since these two contributions cancel each other by definition.}

The one-loop matching is done most efficiently by using the method of regions~\cite{Manohar:1997qy,Beneke:1997zp}: the hard part of the diagrams in the UV theory directly gives the matching contribution, up to finite renormalizations. Since the Fierz-evanescent operator $\E_{qqq\ell}^{(F)}$ receives a tree-level matching contribution, the finite renormalizations that compensate its insertions into one-loop Green's functions need to be taken into account explicitly in the determination of the one-loop matching contribution. The reason is that upon expansion of the integrands in all soft scales, the EFT loop diagrams including the effect of evanescent insertions become vanishing scaleless integrals, hence in the matching equations the finite counterterms cannot be neglected when using the method of regions.  In the present case, these terms consist purely of $SU(2)_L$ contributions to $C_{qqq\ell}$ and Yukawa contributions to $C_{duq\ell}$ and $C_{qque}$.

In terms of the \msbar{} parameters of the UV theory, we find that there are no QCD one-loop corrections to the matching. The $SU(2)_L$ loops only affect $C_{qqq\ell}$, whereas $C_{duq\ell}$ and $C_{duue}$ receive $U(1)_Y$ contributions. In particular, the $U(1)_Y$ one-loop matching contribution to $C_{duq\ell}$ has no logarithmic terms but depends on the scheme parameter $a_\mathrm{ev}$. Including the Yukawa contributions, the complete one-loop matching contributions are
\begin{align}
	\Delta C_{\substack{duq\ell\\prst}} &= \frac{\alpha_1}{4\pi} \frac{(1+2a_\mathrm{ev}) (\y_u - \y_d)(\y_\ell-\y_q) Y^{Q,RR\dagger}_{rp}Y^{LL}_{st}}{2M^2}
		 - \frac{1}{12\pi^2M^2} (Y_d^* Y^{Q,LL\dagger} Y_u^\dagger)_{pr} Y^{LL}_{st} \nn
		&\quad + \frac{1}{64\pi^2M^2} \left( 3 + 2 \log\!\left( \frac{\mu^2}{M^2}\right) \right) \begin{aligned}[t]
			& \Bigl[ 2 (Y_d^* Y^{Q,LL\dagger} )_{ps} (Y^{RR\top} Y_e)_{rt} - ( Y^{Q,RR\dagger} Y_d )_{rs} (Y_d^* Y^{LL} )_{pt} \\
			& + (Y^{Q,RR*} Y_u)_{ps} (Y^{RR\top} Y_e)_{rt} - (Y^{Q,RR*} Y_u)_{ps} ( Y_u^* Y^{LL} )_{rt} \Bigr] \end{aligned} \nn
		&\quad - \frac{1}{96\pi^2M^2} \left( 5 + 6 \log\!\left( \frac{\mu^2}{M^2}\right) \right) \Bigl[ (Y_d^* Y^{Q,LL\dagger})_{ps} (Y_u^* Y^{LL})_{rt} + (Y_u^* Y^{Q,LL\dagger})_{rs} (Y_d^* Y^{LL})_{pt} \Bigr] \nn
		&\quad + \frac{1}{128\pi^2M^2} \left( 7 + 6 \log\!\left( \frac{\mu^2}{M^2}\right) \right) \begin{aligned}[t]
			& \Bigl[ ( Y^{Q,RR*} Y^{RR\dagger} Y^{RR} )_{pr} Y^{LL}_{st} \\
			& + 8 Y^{Q,RR\dagger}_{rp} (Y^{Q,LL\dagger} Y^{Q,LL} Y^{LL} )_{st} \Bigr] \end{aligned} \nn
		&\quad - \frac{1}{32\pi^2M^2} \left( 1 + 2 \log\!\left( \frac{\mu^2}{M^2}\right) \right) \begin{aligned}[t]
			& \Bigl[ ( Y^{Q,RR\dagger} Y^{Q,RR} Y^{Q,RR\dagger} )_{rp} Y^{LL}_{st} \\
			& + Y^{Q,RR\dagger}_{rp} (Y^{LL} Y^{LL\dagger} Y^{LL} )_{st} \Bigr] \, , \end{aligned} \nn
	\Delta C_{\substack{qque\\prst}} &= 
		 \frac{1}{24\pi^2M^2} Y^{Q,LL\dagger}_{pr} (Y_u^* Y^{LL} Y_e^\dagger )_{st} \nn
		&\quad - \frac{1}{128\pi^2M^2} \left( 3 + 2 \log\!\left( \frac{\mu^2}{M^2}\right) \right) \begin{aligned}[t]
			& \Bigl[  2 (Y^{Q,LL\dagger} Y_u^\dagger )_{ps} (Y^{RR} Y_u)_{tr} + 2 (Y^{Q,LL\dagger} Y_u^\dagger )_{rs} (Y^{RR} Y_u)_{tp} \\
			& - (Y^{Q,RR\dagger} Y_d)_{sp} (Y^{LL} Y_e^\dagger)_{rt} - (Y^{Q,RR\dagger} Y_d)_{sr} (Y^{LL} Y_e^\dagger)_{pt} \\
			&  + (Y^{Q,RR\dagger} Y_d)_{sp} (Y^{RR} Y_u)_{tr} + (Y^{Q,RR\dagger} Y_d)_{sr} (Y^{RR} Y_u)_{tp}
			 \Bigr] \end{aligned} \nn
		&\quad + \frac{1}{192\pi^2M^2} \left( 5 + 6 \log\!\left( \frac{\mu^2}{M^2}\right) \right) \Bigl[ (Y^{Q,LL\dagger} Y_u^\dagger )_{ps} (Y^{LL} Y_e^\dagger )_{rt} + (Y^{Q,LL\dagger} Y_u^\dagger )_{rs} (Y^{LL} Y_e^\dagger )_{pt} \Bigr] \nn
		&\quad + \frac{1}{128\pi^2M^2} \left( 7 + 6 \log\!\left( \frac{\mu^2}{M^2}\right) \right) \begin{aligned}[t]
			& \Bigl[ ( Y^{LL} Y^{LL\dagger} Y^{Q,LL\dagger} )_{pr} Y^{RR}_{ts} + ( Y^{LL} Y^{LL\dagger} Y^{Q,LL\dagger} )_{rp} Y^{RR}_{ts} \\
			& + 2 Y^{Q,LL\dagger}_{pr} (Y^{Q,RR\dagger} Y^{Q,RR} Y^{RR\top} )_{st} \Bigr] \end{aligned} \nn
		&\quad - \frac{1}{32\pi^2M^2} \left( 1 + 2 \log\!\left( \frac{\mu^2}{M^2}\right) \right) \begin{aligned}[t]
			& \Bigl[ 4 ( Y^{Q,LL\dagger} Y^{Q,LL} Y^{Q,LL\dagger} )_{pr} Y^{RR}_{ts} \\
			& + Y^{Q,LL\dagger}_{pr} (Y^{RR} Y^{RR\dagger} Y^{RR} )_{ts} \Bigr] \, , \end{aligned} \nn
	\Delta C_{\substack{qqq\ell\\prst}} &= \frac{\alpha_2}{4\pi} \frac{1}{M^2} \biggl[ \begin{aligned}[t] 
		& \left(3+2\log\!\left(\frac{\mu^2}{M^2}\right) \right) Y^{Q,LL\dagger}_{pr} Y^{LL}_{st} + \frac{4}{3}\left(1+3\log\!\left(\frac{\mu^2}{M^2}\right) \right) Y^{Q,LL\dagger}_{ps} Y^{LL}_{rt} \\
		&\quad + {\frac{14}{3} Y^{Q,LL\dagger}_{rs}Y^{LL}_{pt}} \biggr] 
        - \frac{1}{192\pi^2 M^2} \left( 7 + 6 \log\!\left( \frac{\mu^2}{M^2} \right) \right) \times \end{aligned} \nn*
		&\quad \begin{aligned}[t]
			\times & \biggl[ ( Y^{LL} Y^{LL\dagger} Y^{Q,LL\dagger} )_{pr} Y^{LL}_{st} + ( Y^{LL} Y^{LL\dagger} Y^{Q,LL\dagger} )_{rp} Y^{LL}_{st} \\
			& + 2 ( Y^{LL} Y^{LL\dagger} Y^{Q,LL\dagger} )_{rs} Y^{LL}_{pt} + 2 ( Y^{LL} Y^{LL\dagger} Y^{Q,LL\dagger} )_{sr} Y^{LL}_{pt} \\
			& + 8 Y^{Q,LL\dagger}_{pr} ( Y^{Q,LL\dagger} Y^{Q,LL} Y^{LL} )_{st}  + 16 Y^{Q,LL\dagger}_{rs} ( Y^{Q,LL\dagger} Y^{Q,LL} Y^{LL} )_{pt} \biggr] \end{aligned} \nn
	        &\quad + \frac{1}{48\pi^2 M^2} \left( 1 + 2 \log\!\left( \frac{\mu^2}{M^2} \right) \right) \begin{aligned}[t]
			\times & \biggl[ Y^{Q,LL\dagger}_{pr} ( Y^{LL} Y^{LL\dagger} Y^{LL} )_{st} + 2 Y^{Q,LL\dagger}_{rs} ( Y^{LL} Y^{LL\dagger} Y^{LL} )_{pt} \\
			& + 4 ( Y^{Q,LL\dagger} Y^{Q,LL} Y^{Q,LL\dagger} )_{pr} Y^{LL}_{st} \\
			& + 8 ( Y^{Q,LL\dagger} Y^{Q,LL} Y^{Q,LL\dagger} )_{rs} Y^{LL}_{pt} \biggr] \, , \end{aligned} \nn
	\Delta C_{\substack{duue\\prst}} &= \frac{\alpha_1}{4\pi} \frac{(\y_u - \y_d)(\y_e-\y_u)}{M^2} \left(3 + 2\log\!\left( \frac{\mu^2}{M^2} \right) \right) \biggl[ Y^{Q,RR\dagger}_{sp} Y^{RR}_{tr} - \frac{1}{2} Y^{Q,RR\dagger}_{rp} Y^{RR}_{ts} \biggr] \nn
		&\quad + \frac{1}{128\pi^2M^2} \left( 7 + 6 \log\!\left( \frac{\mu^2}{M^2}\right) \right) \begin{aligned}[t]
			& \Bigl[ ( Y^{Q,RR*} Y^{RR\dagger} Y^{RR} )_{pr} Y^{RR}_{ts} \\
			& + 2 Y^{Q,RR\dagger}_{rp} (Y^{Q,RR\dagger} Y^{Q,RR} Y^{RR\top} )_{st} \Bigr] \end{aligned} \nn
		&\quad - \frac{1}{32\pi^2M^2} \left( 1 + 2 \log\!\left( \frac{\mu^2}{M^2}\right) \right) \begin{aligned}[t]
			& \Bigl[ ( Y^{Q,RR\dagger} Y^{Q,RR} Y^{Q,RR\dagger} )_{rp} Y^{RR}_{ts} \\
			& + Y^{Q,RR\dagger}_{rp} (Y^{RR} Y^{RR\dagger} Y^{RR} )_{ts} \Bigr] \, , \end{aligned}
\end{align}
where $\alpha_2 = g_2^2/(4\pi)$, $(\cdot)^\top$ denotes transposition in flavor space, and $(\cdot)^\dagger = (\cdot)^{*\top}$.

\subsection{Cancellation of scheme and scale dependences}

In the following, we check the cancellation of scheme and scale dependences between the two-loop RGEs and the one-loop matching. We start by defining the input parameters in the \msbar{} scheme at the fixed scale of the LQ mass $\mu=M$. These are the \msbar{} LQ mass $M(M)$ itself and the \msbar{} LQ Yukawa couplings $Y^{LL}(M)$, $Y^{RR}(M)$, $Y^{Q,LL}(M)$, and $Y^{Q,RR}(M)$. We then relate the \msbar{} parameters at $\mu=M$ to the ones at the matching scale $\mu = \mu_M$, to NLO or LL accuracy. Since $\mu_M \approx M$, this relation does not involve large logarithms and the difference between a fixed-order NLO relation and a resummation is immaterial. At one loop, the scale dependence of the \msbar{} LQ mass is
\begin{equation}
	[ \dot M ]_1 = M \begin{aligned}[t] & \Big( - 3 g_1^2 y_{S_1}^2 - 4 g_3^2 + \< Y^{RR} Y^{RR\dagger}\> + 2 \< Y^{LL} Y^{LL\dagger} \> \\
				&\quad + 2 \< Y^{Q,RR} Y^{Q,RR\dagger} \>  + 8 \< Y^{Q,LL} Y^{Q,LL\dagger} \> \Big) \, , \end{aligned}
\end{equation}
whereas for the running of the LQ Yukawa couplings we find
\begin{align}
	[ \dot Y^{LL} ]_1 &= - \left( \frac{5}{6} g_1^2 + \frac{9}{2} g_2^2 + 4 g_3^2 \right) Y^{LL} \nn
		&\quad + 2 Y_u^\top Y^{RR\top} Y_e + \frac{1}{2} Y^{LL} Y_e^\dagger Y_e + \frac{1}{2} Y_u^\top Y_u^* Y^{LL} + \frac{1}{2} Y_d^\top Y_d^* Y^{LL} \nn
		&\quad + 2 Y^{LL} Y^{LL\dagger} Y^{LL} - 12 Y^{Q,LL\dagger} Y^{Q,LL} Y^{LL} \nn
		&\quad + Y^{LL} \left( \< Y^{RR} Y^{RR\dagger} \> + 2\<Y^{LL} Y^{LL\dagger}\> + 2 \< Y^{Q,RR} Y^{Q,RR\dagger}\> + 8 \< Y^{Q,LL} Y^{Q,LL\dagger} \>
        \right) \, , \nn{}
	[ \dot Y^{RR} ]_1 &= - \left( \frac{13}{3} g_1^2 + 4 g_3^2 \right) Y^{RR} + 4 Y_e^* Y^{LL\top} Y_u^\dagger + Y^{RR} Y_u Y_u^\dagger + Y_e^* Y_e^\top Y^{RR} \nn
		&\quad + 2 Y^{RR} Y^{RR\dagger} Y^{RR} - 3 Y^{RR} Y^{Q,RR\top} Y^{Q,RR*} \nn
		&\quad + Y^{RR} \left( \< Y^{RR} Y^{RR\dagger} \> + 2\<Y^{LL} Y^{LL\dagger}\> + 2 \< Y^{Q,RR} Y^{Q,RR\dagger}\> + 8 \< Y^{Q,LL} Y^{Q,LL\dagger} \>
        \right) \, , \nn{}
	[ \dot Y^{Q,LL} ]_1 &= - \left( \frac{1}{6} g_1^2 + \frac{9}{2} g_2^2 + 8 g_3^2 \right) Y^{Q,LL}  - Y_d^\dagger Y^{Q,RR} Y_u^*  - Y_u^\dagger Y^{Q,RR\top} Y_d^* \nn
		&\quad + \frac{1}{2} Y^{Q,LL} Y_u^\top Y_u^* + \frac{1}{2} Y^{Q,LL} Y_d^\top Y_d^* + \frac{1}{2} Y_u^\dagger Y_u Y^{Q,LL} + \frac{1}{2} Y_d^\dagger Y_d Y^{Q,LL} \nn
		&\quad + 8 Y^{Q,LL} Y^{Q,LL\dagger} Y^{Q,LL} - \frac{3}{2} Y^{Q,LL} Y^{LL} Y^{LL\dagger} - \frac{3}{2} Y^{LL*} Y^{LL\top} Y^{Q,LL} \nn
		&\quad + Y^{Q,LL} \left( \< Y^{RR} Y^{RR\dagger} \> + 2\<Y^{LL} Y^{LL\dagger}\> + 2 \< Y^{Q,RR} Y^{Q,RR\dagger}\> + 8 \< Y^{Q,LL} Y^{Q,LL\dagger} \>
        \right) \, , \nn{}
	[ \dot Y^{Q,RR} ]_1 &= - \left( \frac{5}{3} g_1^2 + 8 g_3^2 \right) Y^{Q,RR} - 8 Y_d Y^{Q,LL} Y_u^\top + Y^{Q,RR} Y_u^* Y_u^\top + Y_d Y_d^\dagger Y^{Q,RR} \nn*
		&\quad + 2 Y^{Q,RR} Y^{Q,RR\dagger} Y^{Q,RR} - \frac{3}{2} Y^{Q,RR} Y^{RR\top} Y^{RR*} \nn*
		&\quad + Y^{Q,RR} \left( \< Y^{RR} Y^{RR\dagger} \> + 2\<Y^{LL} Y^{LL\dagger}\> + 2 \< Y^{Q,RR} Y^{Q,RR\dagger}\> + 8 \< Y^{Q,LL} Y^{Q,LL\dagger} \>
        \right) \, .
\end{align}
The corrections due to higher powers of the LQ Yukawa couplings are included for completeness but are not needed for the check of scheme- and scale-dependence cancellation.

Next, we determine the SMEFT Wilson coefficients at the scale $\mu = \mu_M$ using the one-loop matching. We subsequently apply the two-loop SMEFT RGEs to determine the Wilson coefficients at some low scale $\mu_L$, resumming large logarithms to NLL accuracy. The dependence on the matching scale is determined from the variation in the interval $\mu_M \in [ M/2, 2 M ]$.

For a phenomenological application, the renormalized \msbar{} input parameters should be related to physical observables. In the case of the LQ mass, the relation between \msbar{} mass $M$ at the scale $\mu$ and the pole mass $M_\mathrm{OS}$ is given by,
\begin{equation}
	\label{eq:MOS}
	M^2_\mathrm{OS} = M^2 \biggl[ 1 \begin{aligned}[t]
		&+ \left( \frac{\alpha_1}{4\pi} \y_{S_1}^2 + \frac{\alpha_s}{4\pi} C_F \right) \left( 7 + 3 \log\!\left(\frac{\mu^2}{M^2}\right) \right) \\
		&- \frac{1}{16\pi^2} \left( \< Y^{RR} Y^{RR\dagger}\> + 2 \< Y^{LL} Y^{LL\dagger} \> + 2 \< Y^{Q,RR} Y^{Q,RR\dagger} \>  + 8 \< Y^{Q,LL} Y^{Q,LL\dagger} \> \right) \\
		&\qquad\qquad \times \left( 2 + \log\!\left( -\frac{\mu^2}{M^2} \right) \right) 
        \biggr] \, , \end{aligned}
\end{equation}
where $C_F = 4/3$, $\alpha_1 = g_1^2 / (4\pi)$, $\alpha_s = g_3^2 / (4\pi)$, 
$\< \cdot \>$ denotes the trace in flavor space,
and we neglect the fermion masses, i.e., we compute the pole mass in the unbroken phase. Note that due to the fermionic decay channels, the LQ acquires a width, i.e., the pole mass gets an imaginary part. We again included the effects due to higher powers of the LQ Yukawa couplings just for completeness.
Similarly to the mass, the LQ Yukawa couplings should be related to the physical LQ decay widths at NLO, i.e., including virtual and real corrections.

For the illustration of the cancellation of scheme and matching-scale dependences, these relations to observables are not needed and we simply work with \msbar{} input parameters at the fixed scale $\mu=M$. For simplicity, we restrict the discussion to the gauge interactions, but with the full two-loop RGE results provided as supplementary material, the analysis can be extended to the Yukawa interactions.
Using the flavor decompositions~\eqref{eq:CqqqlDecomposition}, which have diagonal gauge corrections, we can conveniently express the results in terms of the tree-level expressions Eqs.~\eqref{eq:TreeLevelMatching} and~\eqref{eq:TreeLevelMatching2} with renormalized Yukawa couplings $Y(M)$ and \msbar{} mass $M(M)$.

For an analytic check of the matching-scale cancellation, we expand the solution of the RGEs in powers of the \msbar{} gauge couplings at $\mu=M$. At LL accuracy, re-expanded to NLO, this leads to
\begin{align}
	\label{eq:LLMatching}
	\frac{C_{duq\ell}(\mu_L)}{C_{duq\ell}^\mathrm{tree}} \Big|^\mathrm{LL}_\mathrm{NLO} &= 1 - \left( \frac{11\alpha_1}{24\pi} + \frac{9\alpha_2}{8\pi} + \frac{\alpha_s}{\pi} \right) \log\!\left( \frac{\mu_L}{\mu_M} \right) + \ldots \, , \nn
	\frac{C_{qque}(\mu_L)}{C_{qque}^\mathrm{tree}} \Big|^\mathrm{LL}_\mathrm{NLO} &= 1 - \left( \frac{23\alpha_1}{24\pi} + \frac{9\alpha_2}{8\pi} + \frac{\alpha_s}{\pi} \right) \log\!\left( \frac{\mu_L}{\mu_M} \right) + \ldots \, , \nn
	\frac{C_{duue}^{(+)}(\mu_L)}{C_{duue}^{(+),\mathrm{tree}}} \Big|^\mathrm{LL}_\mathrm{NLO} &= 1 - \left( \frac{13\alpha_1}{6\pi} + \frac{\alpha_s}{\pi} \right) \log\!\left( \frac{\mu_L}{\mu_M} \right) + \ldots \, , \nn
	\frac{C_{duue}^{(-)}(\mu_L)}{C_{duue}^{(-),\mathrm{tree}}} \Big|^\mathrm{LL}_\mathrm{NLO} &= 1 - \left( -\frac{7\alpha_1}{6\pi} + \frac{\alpha_s}{\pi} \right) \log\!\left( \frac{\mu_L}{\mu_M} \right) + \ldots \, , \nn
	\frac{S_{qqq\ell}(\mu_L)}{S_{qqq\ell}^\mathrm{tree}} \Big|^\mathrm{LL}_\mathrm{NLO} &=  1 - \left( \frac{\alpha_1}{12\pi} + \frac{15\alpha_2}{4\pi} + \frac{\alpha_s}{\pi} \right) \log\!\left( \frac{\mu_L}{\mu_M} \right) + \ldots \,   , \nn
	A_{qqq\ell}(\mu_L) \Big|^\mathrm{LL}_\mathrm{NLO} &= \left[ 1 - \left( \frac{\alpha_1}{12\pi} - \frac{9 \alpha_2}{4\pi} + \frac{\alpha_s}{\pi} \right) \log\!\left( \frac{\mu_L}{\mu_M} \right) + \ldots \, \right] A_{qqq\ell}^\mathrm{tree} , \nn
	\frac{M_{qqq\ell}(\mu_L)}{M_{qqq\ell}^\mathrm{tree}} \Big|^\mathrm{LL}_\mathrm{NLO} &= 1 - \left( \frac{\alpha_1}{12\pi} + \frac{3\alpha_2}{4\pi} + \frac{\alpha_s}{\pi} \right) \log\!\left( \frac{\mu_L}{\mu_M} \right) + \ldots \, ,
\end{align}
where the ellipses denote neglected Yukawa terms. The dependence on the matching scale $\mu_M$ appears already at $\O(\alpha)$ when using tree-level matching and one-loop running. In Eq.~\eqref{eq:LLMatching}, we include the re-expanded RGE contribution for $A_{qqq\ell}$, although in the present case of the LQ model, the matching gives $A_{qqq\ell}^\mathrm{tree} = 0$.  At NLL accuracy, re-expanded to next-to-next-to-leading order (NNLO), we find, e.g.,
\begin{align}
	\label{eq:CduqlNLLMatched}
	\frac{C_{duq\ell}(\mu_L)}{C_{duq\ell}^\mathrm{tree}} \Big|^\mathrm{NLL}_\mathrm{NNLO} &= 1 - \left( \frac{11\alpha_1}{24\pi} + \frac{9\alpha_2}{8\pi} + \frac{\alpha_s}{\pi} \right) \log\!\left( \frac{\mu_L}{M} \right) \nn*
		&\quad + \left( -\frac{781\alpha_1^2}{1152\pi^2} + \frac{195\alpha_2^2}{128\pi^2} + \frac{9\alpha_s^2}{4\pi^2} + \frac{33\alpha_1\alpha_2}{64\pi^2} + \frac{11\alpha_1\alpha_s}{24\pi^2} + \frac{9\alpha_2\alpha_s}{8\pi^2} \right) \log^2\!\left( \frac{\mu_L}{M} \right) \nn*
		&\quad - \frac{(1+2a_\mathrm{ev})\alpha_1}{12\pi} \nn
		&\quad + \bigg( \begin{aligned}[t]
			&\frac{(811-1136a_\mathrm{ev})\alpha_1^2}{2304\pi^2} - \frac{541\alpha_2^2}{256\pi^2} - \frac{31\alpha_s^2}{24\pi^2} \\
			& + \frac{3(5+8a_\mathrm{ev})\alpha_1\alpha_2}{128\pi^2} + \frac{(37+24a_\mathrm{ev})\alpha_1\alpha_s}{144\pi^2} + \frac{9\alpha_2\alpha_s}{16\pi^2} \bigg) \log\!\left( \frac{\mu_L}{M} \right) \end{aligned} \nn
		&\quad \color{gray} + \left( -\frac{1379\alpha_1^2}{2304\pi^2}
				+ \frac{541\alpha_2^2}{256\pi^2} 
				+ \frac{31\alpha_s^2}{24\pi^2} 
				- \frac{3\alpha_1\alpha_2}{128\pi^2} 
				- \frac{25\alpha_1\alpha_s}{144\pi^2} 
				- \frac{9\alpha_2\alpha_s}{16\pi^2} \right) \log\!\left( \frac{\mu_M}{M} \right) \, , 
\end{align}
where the couplings are evaluated at $\mu=M$. The NLL result allows us to make a few observations. The first two lines in Eq.~\eqref{eq:CduqlNLLMatched} contain the LL terms, while the next three lines are NLL terms (both truncated at NNLO). While the leading logs are scheme independent, the NLL terms contain scheme dependences, in particular a dependence on $a_\mathrm{ev}$. The reason is that the Wilson coefficient itself is a scheme-dependent quantity beyond tree-level. The last line contains a logarithm of the matching scale over the LQ mass, which is not a large logarithm. These terms (in gray) are beyond the accuracy of our calculation, as we did not compute the renormalization of the LQ model at two loops. We find that the dependence on the matching scale $\mu_M$ is shifted to NNLO as expected, i.e., all NLO matching-scale dependences cancel. Regarding the scheme dependence, we find that the matching-scale-dependent terms do not depend on $a_\mathrm{ev}$: we observe a cancellation of such $a_\mathrm{ev}$-dependent terms between the two-loop RGE and the one-loop matching. The remaining $a_\mathrm{ev}$-dependence in Eq.~\eqref{eq:CduqlNLLMatched} will cancel instead at the low scale $\mu_L$, when combining the two-loop RGE with a NLO matrix-element or matching calculation. The same structure is found for the NLL results of the other Wilson coefficients, which are included in the supplementary material.

\begin{figure}[t]
	\centering
	\hspace{1cm}\scalebox{0.8}{\input{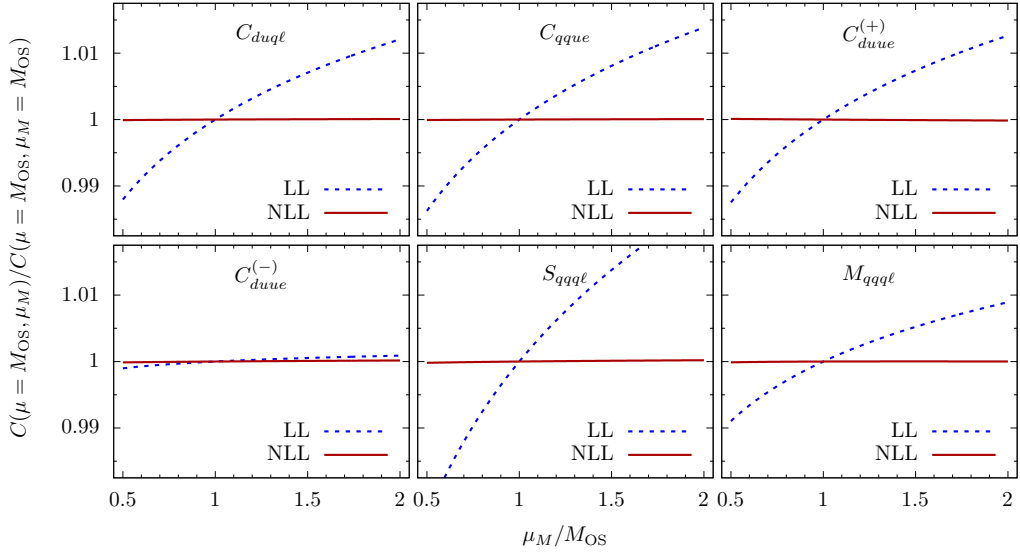}} \\[1cm]
	\caption{Comparison of the LL and NLL dependence of the Wilson coefficients on the matching scale $\mu_M$, normalized to a matching scale of $\mu_M = M$, which is chosen at the unification scale, $M = 6.5 \times 10^{15} \GeV$. We only include the contribution of the gauge interactions.}
	\label{fig:ScaleDependence}
\end{figure}

The reduction of the matching-scale dependence from NLO to NNLO is shown in Fig.~\ref{fig:ScaleDependence}, where we display the Wilson coefficients as a function of the matching scale, varied in the interval $\mu_M \in [ M/2, 2 M ]$ and normalized to the value at $\mu_M = M$. In the case of $C_{duue}^{(-)}$, the small matching-scale dependence at LL results from an accidental numerical cancellation between $SU(3)_c$ and $U(1)_Y$ corrections, see Eq.~\eqref{eq:LLMatching}.

\begin{figure}[t]
	\centering
	\hspace{1cm}\scalebox{0.8}{\input{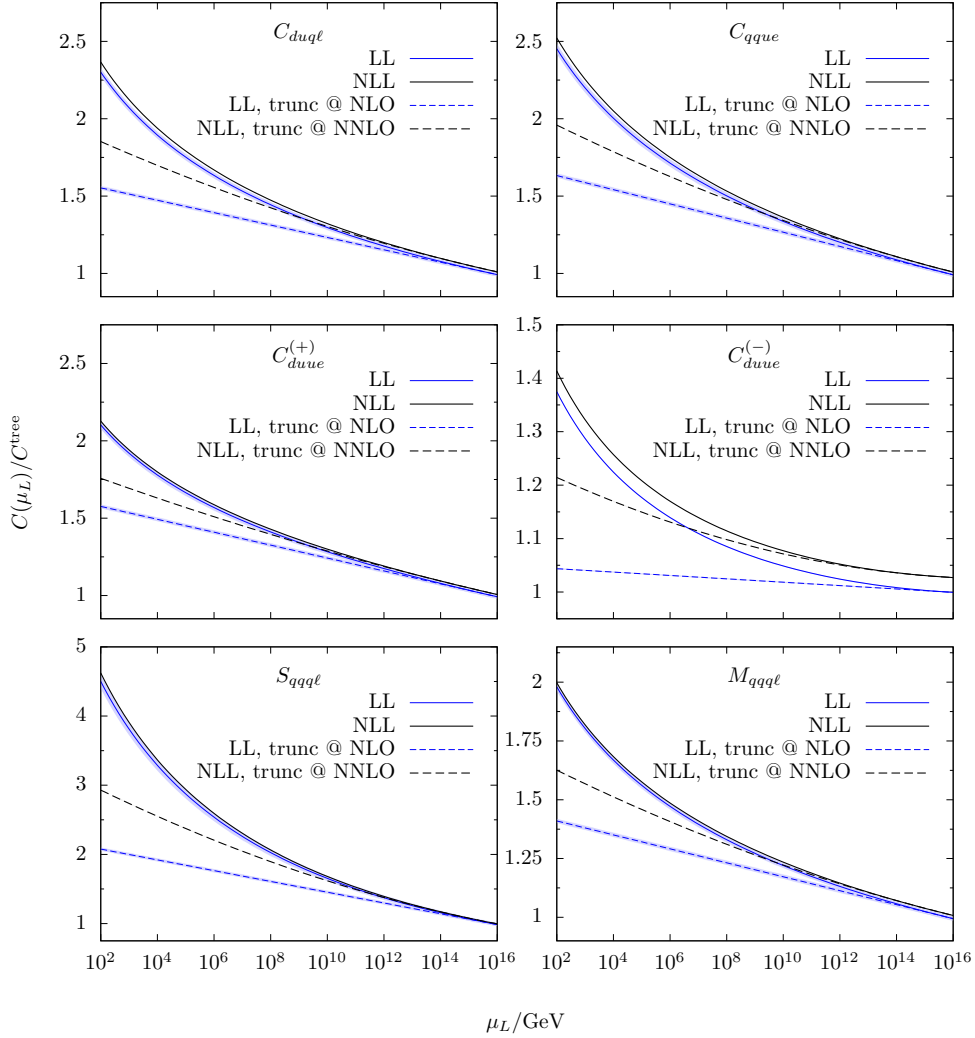}} \\[1cm]
	\caption{Comparison of the LL and NLL running of the Wilson coefficients due to the gauge interactions, normalized to the tree-level expression. The uncertainties are obtained by varying the matching scale between $\mu_M = M/2$ and $\mu_M = 2 M$. The solid lines include a resummation of large logarithms, while the dashed lines are the re-expanded LL and NLL results, truncated at NLO and NNLO, respectively.}
	\label{fig:Running}
\end{figure}

In Fig.~\ref{fig:Running}, we show the running of the Wilson coefficients due to the gauge interactions. The solid lines depict the numerical solutions of the RGEs, i.e., a full resummation of leading and next-to-leading logarithms. The dashed lines show a perturbative expansion of LL and NLL results to NLO and NNLO, respectively, as in Eqs.~\eqref{eq:LLMatching} and~\eqref{eq:CduqlNLLMatched}. As expected, for a large scale separation the resummation of large logarithms is indispensable. The NLL corrections are mostly of the order of the naive LL uncertainty estimate obtained from the matching-scale variation, with the exception of $C_{duue}^{(-)}$, where the scale variation significantly underestimates the NLL corrections due to the aforementioned numerical cancellation of $SU(3)_c$ and $U(1)_Y$ contributions. The NLL uncertainties estimated with the variation of the matching scale are so small that the error bands are not visible at the scale of the plots. Additional finite one-loop corrections arise when relating the \msbar{} parameters to observables, which are not included in Fig.~\ref{fig:Running}, see Eq.~\eqref{eq:MOS}. Overall, the NLL corrections are surprisingly small, but we stress that in Fig.~\ref{fig:Running} we only show the evolution down to the electroweak scale. If we just consider QCD corrections, the single-gauge-coupling solution of the diagonalized NLL RGE has the simple form
\begin{equation}
	C(\mu_L) = \left(1 + \frac{\alpha_s(\mu_L)}{4\pi} J \right) \left( \frac{\alpha_s(\mu_M)}{\alpha_s(\mu_L)} \right)^{\frac{\gamma_0}{2 b_0^{g_3}}} \left(1 - \frac{\alpha_s(\mu_M)}{4\pi} J \right) C(\mu_M) \, , \quad J = \frac{\gamma_0 b_1^{g_3}}{2 (b_0^{g_3})^2} - \frac{\gamma_1}{2 b_0^{g_3}} \, .
\end{equation}
In the case of $C_{duq\ell}$, we have $\gamma_0 = -4$, $\gamma_1 = -62/3$, hence with the coefficients of the beta function $b_0^{g_3} = 7$ and $b_1^{g_3} = 26$, we find $J=61/147$. With $\alpha_s(\mu_L) = 0.118$, this leads to a small (sub-percent level) one-loop correction even at the electroweak scale. We expect to find larger NLL effects when evolving down to the hadronic scale using the two-loop RGE in the LEFT~\cite{Naterop:2025lzc}. New effects will also arise from the Yukawa contributions~\cite{Hou:2005iu}, e.g., $A_{qqq\ell}$ receives mixing contributions from the other Wilson coefficients starting at two loops.


\section{Conclusions}
\label{sec:Conclusions}

Baryon-number conservation, observed experimentally, e.g., via the stability of the proton, provides one of the most stringent limits on extensions of the SM. In fact, it requires the scale of grand unification to be at least $10^{15}$\,GeV. Therefore, in such theories, the separation between the BNV scale and the experimental scale (i.e., the nuclear scale) is huge. In order to obtain accurate predictions, the use of effective field theories is necessary to resum the resulting large logarithms.

Model independently, BNV operators first appear at the dimension-6 level, where four different $SU(2)_L$ invariant types exist in the SMEFT (not counting flavor indices). In this article, we calculated the two-loop renormalization-group evolution of these operators, including the gauge interactions ($SU(3)_c$, $SU(2)_L$, and $U(1)_Y$) and the Higgs Yukawa interactions. We also checked that the Higgs self-interactions do not contribute to the two-loop RGEs. Together with the known one-loop matching of the SMEFT on LEFT and the two-loop RGEs of LEFT, this can be used to calculate BNV processes at NLL accuracy, provided that the matching of the UV theory under investigation and the matching to a non-perturbative determination of the hadronic matrix elements are available at the same level of accuracy.

In this context, we calculated the one-loop matching of the $S_1$ leptoquark onto the SMEFT. Interestingly, this leptoquark generates all four gauge-invariant dimension-6 SMEFT operators at tree level, such that one can check the cancellation of the leading matching-scale dependence as well as of the scheme choice with the simple UV model. We find that the scale dependence is in fact significantly reduced, providing at the same time an additional check of the correctness of our calculation.

A detailed numerical analysis of the full NLL effects in the EFT framework and phenomenological implications will be presented in an upcoming publication~\cite{Banik:2025inprep}.

	
	\section*{Acknowledgements}
	\addcontentsline{toc}{section}{\numberline{}Acknowledgements}

	We thank M.~Hoferichter and A.~V.~Manohar for useful discussions.
	SB is supported by a UZH Postdoc Grant (Grant No.~[FK-24-100]).
	Financial support by the Swiss National Science Foundation (Project Nos.\ PP00P21\_76884 and PCEFP2\_194272) is gratefully acknowledged.

	
	\appendix
	

\section{Conventions}
\label{sec:Conventions}

\subsection{Dirac algebra}

The matrix $\gamma_5$ and the chiral projectors are defined by
\begin{equation}
	\gamma_5 := i \gamma^0 \gamma^1 \gamma^2 \gamma^3 = \frac{i}{4!} \epsilon_{\mu\nu\lambda\sigma} \gamma^\mu \gamma^\nu \gamma^\lambda \gamma^\sigma \, , \quad P_L = \frac{1}{2} ( 1 - \gamma_5 ) \, , \quad P_R = \frac{1}{2} ( 1 + \gamma_5 ) \, ,
\end{equation}
where the Levi-Civita symbol is normalized to $\epsilon_{0123} = +1$.

The charge-conjugation matrix fulfills
\begin{equation}
	\label{eq:ChargeConjugationMatrix}
	C \gamma_\mu C^{-1} = - \gamma_\mu^T \, , \quad C = C^* = - C^{-1} = - C^\dagger = - C^T \, ,
\end{equation}
which in 4 space-time dimensions is realized by $C = i \gamma^2 \gamma^0$. We take the relations~\eqref{eq:ChargeConjugationMatrix} to be true also in $D=4-2\varepsilon$ space-time dimensions~\cite{Belusca-Maito:2020ala}.

\subsection[$SU(N)$ algebra]{\boldmath $SU(N)$ algebra}

For $SU(3)_c$, we use Hermitian generators
\begin{equation}
	T^A = \frac{\lambda^A}{2} \, , \quad \tr[T^A T^B] = \frac{1}{2} \delta^{AB} \, ,
\end{equation}
where $\lambda^A$ are the Gell-Mann matrices. The quadratic Casimir operators in the fundamental and adjoint representations are
\begin{equation}
	\label{eq:SUNCasimir}
	T^A_{\alpha\beta} T^A_{\beta\gamma} = C_F \delta_{\alpha\gamma} = \frac{N_c^2 - 1}{2N_c} \delta_{\alpha\gamma} \, , \quad f^{ABC} f^{ABD} = C_A \delta^{CD} = N_c \delta^{CD} \, .
\end{equation}

For $SU(2)_L$, the generators are
\begin{equation}
	t^I = \frac{\tau^I}{2} \, , \quad \tr[t^I t^J] = \frac{1}{2} \delta^{IJ} \, ,
\end{equation}
where $\tau^I$ are the Pauli matrices.

\subsection{Conventions for the supplementary material}

\begin{table}[t]
	\centering
	\small
	\begin{tabular}{lllc}
		\toprule
		variable							& code name				& explanation						\\
		\midrule
		\midrule
		$n_e$							& \verb$nf[e]$				& number of right-chiral electron flavors \\
		$n_u$							& \verb$nf[u]$				& number of right-chiral up-type quark flavors \\
		$n_d$							& \verb$nf[d]$				& number of right-chiral down-type quark flavors \\
		$n_\ell$							& \verb$nf[l]$				& number of left-chiral lepton doublet flavors \\
		$n_q$							& \verb$nf[q]$				& number of left-chiral quark doublet flavors \\
		\midrule
		$b_{0}^{g_1}$						& \verb$b0g1$				& coefficient of one-loop $U(1)_Y$ $\beta$-function \\
		$b_{0}^{g_2}$						& \verb$b0g2$				& coefficient of one-loop $SU(2)_L$ $\beta$-function \\
		$b_{0}^{g_3}$						& \verb$b0g3$				& coefficient of one-loop $SU(3)_c$ $\beta$-function \\
		\midrule
		$\y_e = -1$						& \verb$y[e]$				& right-chiral electron hypercharge \\
		$\y_u = 2/3$						& \verb$y[u]$				& right-chiral up-quark hypercharge \\
		$\y_d = -1/3$						& \verb$y[d]$				& right-chiral down-quark hypercharge \\
		$\y_\ell = -1/2$						& \verb$y[l]$				& left-chiral lepton doublet hypercharge \\
		$\y_q = 1/6$						& \verb$y[q]$				& left-chiral quark doublet hypercharge \\
		$\y_H = 1/2$						& \verb$y[H]$				& Higgs-doublet hypercharge \\
		\midrule
		$g_1$, $g_2$, $g_3$				& \verb$g1$, \verb$g2$, \verb$g3$	& gauge couplings \\
		\midrule
		$Y_e, Y_e^\dagger$					& \verb$Y[e]$, \verb$Ydag[e]$		& electron Yukawa matrix \\
		$Y_u, Y_u^\dagger$					& \verb$Y[u]$, \verb$Ydag[u]$		& up-quark Yukawa matrix \\
		$Y_d, Y_d^\dagger$					& \verb$Y[d]$, \verb$Ydag[d]$		& down-quark Yukawa matrix \\
		\midrule
		$M$								& \verb$MS1$				& $S_1$ leptoquark \msbar{} mass \\
		\midrule
		$Y^{RR}, Y^{RR\dagger}$				& \verb$YRR$, \verb$YRRdag$		& $S_1$ leptoquark Yukawa couplings \\
		$Y^{LL}, Y^{LL\dagger}$				& \verb$YLL$, \verb$YLLdag$			&  \\
		$Y^{Q,RR}, Y^{Q,RR\dagger}$			& \verb$YQRR$, \verb$YQRRdag$		&  \\
		$Y^{Q,LL}, Y^{Q,LL\dagger}$			& \verb$YQLL$, \verb$YQLLdag$		&  \\
		\midrule
		$\< A \cdots B \>$					& \verb$flTr[A,...,B]$			& trace in flavor space \\
		$(A \cdots B)_{pr}$					& \verb$FCHN[{A,...,B},p,r]$	& flavor chain: element $p,r$ of a product \\
																&&  of flavor-space matrices \\
		$A^\top$							& \verb$transp[A]$			& matrix transposition in flavor space \\
		\midrule
		$C_{duq\ell}$, $C_{qque}$,				& \verb$Cduql$, \verb$Cqque$,			& Wilson coefficients \\
		$C_{qqq\ell}$, $C_{duue}$				& \verb$Cqqql$	, \verb$Cduue$			& \\
		\midrule
		$a_\mathrm{ev}$					& \verb$aev$				& evanescent scheme parameter \\
		\midrule
		$[\dot X ]_2$						& \verb$dot2[X]$			& two-loop contribution to RGE, see Eq.~\eqref{eq:RGENotation} \\
		\bottomrule
	\end{tabular}
	\caption{Variables appearing in the code with the two-loop RGE and one-loop matching results, provided as supplementary material.}
	\label{tab:CodeVariables}
\end{table}

The full results for the two-loop RGEs and the one-loop LQ matching are provided as supplementary material in the form of a \texttt{Mathematica} notebook. The variables used in the notebook are defined in Tab.~\ref{tab:CodeVariables}. We employ conventions analogous to Refs.~\cite{Naterop:2023dek,Naterop:2025cwg}, in particular we use a matrix-style notation in flavor space that avoids sums over dummy indices. Rank-4 flavor tensors are treated using the notation
\begin{equation}
	\mathrm{FFA}(C)_{pr} \mathrm{FFB}(C)_{st} := C_{prst} \, ,
\end{equation}
where the first two flavor indices of the tensor $C$ are attached to the symbol FFA and the last two indices to FFB. This allows us to extend the matrix notation to the Wilson coefficients of four-fermion operators.

	\phantomsection
	\addcontentsline{toc}{section}{\numberline{}References}
	\bibliographystyle{utphysmod}
	\bibliography{Literature}
	
\end{document}